\documentclass[twocolumn]{aastex62}

\hypersetup{linkcolor=red,citecolor=blue,filecolor=cyan,urlcolor=blue}

\usepackage{braket}
\usepackage{amsmath}
\usepackage{bm}

\newcommand{\lya}{Ly$\alpha\,$}
\newcommand{\lyb}{Ly$\beta\,$}
\newcommand{\lyc}{Ly$\gamma\,$}

\newcommand{\cmpch}{cMpc$\,h^{-1}\,$}

\newcommand{\tdr}{temperature-density relation~}

\accepted{\today}
\submitjournal{ApJ}

\shorttitle{Anomalous IGM Opacity in the \lya\ and \lyb\ Forest}
\shortauthors{A.-C. Eilers et al.}

\begin{document}

\title{\textbf{Anomaly in the Opacity of the Post-Reionization Intergalactic Medium in the \lya\ and \lyb\ Forest}}

\author[0000-0003-2895-6218]{Anna-Christina Eilers}
\affiliation{Max-Planck-Institute for Astronomy, K\"onigstuhl 17, 69117 Heidelberg, Germany}
\affiliation{International Max Planck Research School for Astronomy \& Cosmic Physics at the University of Heidelberg}
\author[0000-0002-7054-4332]{Joseph F. Hennawi}
\affiliation{Max-Planck-Institute for Astronomy, K\"onigstuhl 17, 69117 Heidelberg, Germany}
\affiliation{Physics Department, University of California, Santa Barbara, CA 93106-9530, USA}
\author[0000-0003-0821-3644]{Frederick B. Davies}
\affiliation{Physics Department, University of California, Santa Barbara, CA 93106-9530, USA}
\author[0000-0002-8723-1180]{Jose O\~norbe}
\affiliation{Institute for Astronomy, University of Edinburgh, Blackford Hill, EH9 3HJ, Edinburgh, United Kingdom} 

\correspondingauthor{Anna-Christina Eilers}
\email{eilers@mpia.de}

\begin{abstract}
  We measure the intergalactic medium (IGM) opacity in the Ly$\alpha$ as well as in the Ly$\beta$
  forest along $19$ quasar sightlines between $5.5\lesssim z_{\rm abs}\lesssim 6.1$, probing the end stages of the reionization epoch. Owing to its lower oscillator strength the Ly$\beta$
  transition is sensitive to different gas temperatures and densities than Ly$\alpha$,
  providing additional constraints on the ionization and thermal state
  of the IGM. A comparison of our measurements to different inhomogeneous reionization models, derived from post-processing the Nyx cosmological hydrodynamical simulation to include spatial fluctuations in the ultraviolet background (UVB) or the gas temperature field, as well as to a uniform reionization model with varying thermal states of the IGM, leads to two primary
  conclusions: First, we find that including the effects of spectral
  noise is key for a proper data to model comparison. Noise
  effectively reduces the sensitivity to high opacity regions, and
  thus even stronger spatial inhomogeneities are required to match the
  observed scatter in the observations than previously inferred.
  Second, we find that models which come close to 
  reproducing the distribution of Ly$\alpha$ effective optical depths
  nevertheless underpredict the Ly$\beta$ opacity at the same
  spatial locations. The origin of this disagreement is not entirely
  clear but models with an inversion in the temperature-density
  relation of the IGM just after reionization is completed match our measurements best, although they still do not fully capture the observations at $z\gtrsim 5.8$. 
\end{abstract}


\section{Introduction}\label{sec:intro}

The first billion years of our universe are at the forefront of
observational and theoretical cosmological research. During this early
evolutionary epoch the intergalactic medium (IGM) transitioned from a
neutral state following recombination to a highly ionized state due to the
radiation emitted by the first stars, galaxies, and quasars. However,
despite significant progress in recent years the details of this
reionization process, such as its precise timing and morphology,
remain unclear.

The absorption features of neutral hydrogen within the IGM imprinted
on quasar sightlines at $z\gtrsim 6$ have proven to be a valuable
observational tool to constrain the Epoch of Reionization (EoR). 
The evolution of the IGM opacity within the \lya\ forest along quasar
sightlines shows a steep rise around $z\gtrsim 5.5$ as well as an
increased scatter in the measurements \citep{Fan2006, Becker2015,
  Eilers2018a, Bosman2018}, suggesting a qualitative change in the
ionization state of the IGM provoked by a decrease in the ionizing
ultraviolet background (UVB) radiation \citep{Calverley2011,
  WyitheBolton2011, Davies2018a, DayalFerrara2018, Kulkarni2019}.

The largest outliers in the IGM opacity measurements come from the detection of a very long Gunn-Peterson trough in the \lya\ forest along the sightline of $\rm ULAS\,J0148+0600$ extending down to $z\sim 5.5$. These outliers have been the subject of extensive modeling efforts, which provide evidence for either large coherent spatial fluctuations in the UVB \citep{DaviesFurlanetto2016, Daloisio2018}, residual fluctuations in the temperature field  (\citealt{Daloisio2015}, but see \citealt{Keating2018}), the imprint of rare but bright sources of ionizing photons \citep{Chardin2015, Chardin2017}, or ``islands'' of residual neutral gas as low as $z\lesssim 5.5$ due to an extended, inhomogeneous reionization process \citep{Kulkarni2019}. 
Distinguishing these scenarios via the distribution of \lya\ forest opacity alone is challenging \citep{Davies2018b}, although the recent discovery of a large-scale underdensity of Ly$\alpha$-emitting galaxies around the $\rm ULAS\,J0148+0600$ Gunn-Peterson trough seems to suggest that UVB fluctuations may be the culprit \citep{Becker2018}.

In this paper, we explore a different tracer co-spatial with the \lya\ forest, namely the \lyb\ forest. Whereas the overly-sensitive \lya\ transition saturates already for neutral gas fractions of $x_{\rm HI}\gtrsim 10^{-4}$, the $\sim 5$ times lower oscillator strength of the \lyb\ transition makes it more sensitive to gas with a higher neutral fractions and thus provides more stringent constraints on the IGM ionization state (e.g. \citealt{Davies2018a}), as well as on its thermal state \citep{OhFurlanetto2005, Trac2008, FurlanettoOh2009}.

This can be understood when having a closer look at the mean transmitted forest flux and the Gunn-Peterson optical depth $\tau$. Assuming a volume-weighted probability density function for IGM physical conditions
(see e.g. \citet{FaucherGiguere2008a}), we can write mean flux over some spatial interval as
\begin{equation}
\langle F \rangle (z) = \int\limits_0^{\infty}\int\limits_0^{\infty}\int\limits_0^{\infty}d\Delta dT d\Gamma_{\rm HI} P(\Delta, T, \Gamma_{\rm HI} |z)\exp(-\tau_i), 
\end{equation}
where 
\begin{equation}
  \tau_i \propto n_{\rm HI}\propto n_{\rm H}x_{\rm HI} \propto \frac{\Delta^2 T^{-0.7}}{\Gamma_{\rm HI}}.
\end{equation}
is the opacity in either the Ly$\alpha$ or Ly$\beta$ transition, and $\Gamma_{\rm HI}$ is the ionization rate of the UVB.
Owing to difference in their oscillator strengths the two transitions take different distinct `moments' of the distribution $P(\Delta, T, \Gamma_{\rm HI} |z)$ encapsulating IGM physical conditions \citep[see also Fig.~$6$ of][]{FurlanettoOh2009}.

In accordance with the standard paradigm for the thermal state of the post-reionization
photoionized IGM, we expect that the majority of the optically thin gas responsible for the absorption in the \lya\ and \lyb\ forests follows a tight relation between the gas
density $\rho$ and its temperature $T$, the `equation of state', which arises from a balance
between photoheating and the aggregate effect of recombination,
inverse Compton cooling, and adiabatic cooling due to the expansion of
the universe:
\begin{equation}
T = T_0 \left(\frac{\rho}{\rho_0}\right)^{\gamma - 1}, \label{eq:tdr}
\end{equation}
where $T_0$ denotes the temperature at average density $\rho_0$ \citep{Hui1997, McQuinn2015}. A slope with $\gamma < 1$ denotes an inverted temperature-density relation, implying that under-dense voids are hotter than over-dense regions in the IGM, while $\gamma = 1$ represents an isothermal temperature-density relation. One expects to find a fiducial value of $\gamma\approx1.5-1.6$ for the IGM long after any reionization events \citep{Hui1997, McQuinn2015}. 

However, during and immediately after reionization events a flat or inverted temperature-density relation is predicted by several studies using hydrodynamical simulations that model the reionization process self-consistently. \citet{Trac2008} showed that the gas near large overdensities ionizes and heats up earlier than the gas in underdense voids, and hence the IGM temperature is inversely proportional to the reionization redshift. Thus, an inverted equation of state naturally arises at the end of the reionization epoch, although with a large scatter. In their late reionization scenario, in which reionization completes at $z\sim6$, an inverted or isothermal temperature-density relation of the low-density gas in the IGM persists until $z\sim 4$. These results have later been reproduced by \citet{Finlator2018}, who model the EoR with an inhomogeneous UVB radiation field, and find that an isothermal temperature-density relation endures well past the end of the reionization process. 

\citet{FurlanettoOh2009} explore the effects of an inhomogeneous reionization process on the temperature-density relation of the IGM with an analytic model, and also predict an inversion of the equation of state during the EoR. They find a degeneracy between a rapidly evolving UVB and temperature field, and conclude that a wider range of densities and different effective temperatures probed by the higher Lyman-series forests is necessary to set tighter constraints on the thermal state of the IGM. 

While several authors have measured the evolution of \lya\ and \lyb\ opacity and their implications for the neutral gas fraction of the IGM \citep{Lidz2002, Songaila2004, Fan2006}, the correspondence between \lya\ and \lyb\ opacities has not yet been studied in detail. 
Here, we address this matter and measure the IGM opacity in both the Ly$\alpha$ and the Ly$\beta$ forest towards the end of the reionization epoch between $5.5\leq z\leq 6.1$ along $19$ quasar sightlines which have $\rm S/N\gtrsim 10$ per pixel and do not show any broad absorption line signatures. We present our data set and the applied methods to measure the IGM opacity in \S~\ref{sec:methods}, whereas our final measurements are shown in \S~\ref{sec:results}. In \S~\ref{sec:sims} we introduce various models of the physical conditions in the post-reionization IGM, which we obtain by post-processing of the \texttt{Nyx} cosmological hydrodynamical simulation, and conduct a comparison between our results and the predictions from these models. We discuss the implications of our results on the EoR in \S~\ref{sec:discussion}, before summarizing our main findings in \S~\ref{sec:summary}. Throughout this paper we assume a cosmology of $h=0.685$, $\Omega_{\rm m}=0.3$ and $\Omega_{\Lambda}=0.7$, which is consistent within the $1\sigma$ errorbars with \citet{Planck2018}. 

\section{Methods}\label{sec:methods}

\subsection{The Quasar Sample}\label{sec:data}

Our original data sample comprises $34$ quasar spectra at $5.77\leq z_{\rm em}\leq 6.54$ and has been publicly released \citep{Eilers2018a} via the \texttt{igmspec} database \citep{igmspec} and the \texttt{zenodo} platform\footnote{\url{https://doi.org/10.5281/zenodo.1467821}}. Previously, we have used this data set to analyze the sizes of quasar proximity zones \citep{Eilers2017a}, as well as the redshift evolution of the IGM opacity within the \lya\ forest  \citep{Eilers2018a}. 
For the study of the joint \lya\ and \lyb\ forest opacity that we present in this paper we take a subset of $19$ quasar spectra with $\rm S/N \geq 10$ per $10\,\rm km\,s^{-1}$ pixel. All spectra have been observed at optical wavelengths ($4000\rm{\AA}-10000\rm{\AA}$) with the Echellette Spectrograph and Imager \citep[ESI;][]{ESI} at the Keck II Telescope between the years of 2001 to 2016. The data were collected from the Keck Observatory Archive\footnote{\url{https://koa.ipac.caltech.edu/cgi-bin/KOA/nph-KOAlogin}} and complemented with our own observations. All observations used slit widths ranging from $0.75"-1.0"$, resulting in a resolution of $R\approx 4000-5400$. 

The details of all individual observations as well as detailed information about the data reduction process can be found in \citet{Eilers2017a, Eilers2018a}. 
We make further improvements on our data reduction in order to avoid biases in our analysis of the opacities within the \lyb\ forest \textit{relative} to the opacities within the \lya\ forest, that could arise due to potential flux calibration issues or in the process of co-adding individual exposures. Hence we correct all final extracted and co-added spectra with a power-law if necessary to match the observed photometry in the $i$- and $z$-band (see Appendix~\ref{app:photometry} for details on this procedure). 
The properties of all quasars in our data sample are listed in Table~$1$ in \citet{Eilers2018a}. 
\vspace*{.2 cm}

\subsection{Continuum Normalization}
We normalize the quasar spectra by their continuum emission, which we
estimate via a principal component analysis (PCA) reconstruction. The idea of the PCA is to represent each continuum
spectrum by a mean spectrum plus a finite number of weighted principal
component spectra \citep{Suzuki2005, Suzuki2006, Paris2011}. Since the quasar spectra in our data set
experience substantial absorption bluewards of the \lya\ emission line
from intervening neutral hydrogen in IGM, we follow the procedure
suggested by \citet{Paris2011} 
and estimate the coefficients for the
reconstructed continuum emission using only pixels redwards of the
\lya\ line, and then apply a projection matrix to obtain the
coefficients for the continuum spectrum covering the whole spectral
range between $1020\rm{\AA}\leq\lambda_{\rm rest}\leq
1600\rm{\AA}$. Since the PCA components do not cover bluer wavelengths, we extend the estimated continuum to bluer
wavelengths by appending the composite quasar spectrum from
\citet{Shull2012} continuously at $\lambda_{\rm rest} < 1020\rm{\AA}$
\citep[see][for details]{Eilers2018a}.

\begin{figure}[t!]
\centering
\includegraphics[width=.5\textwidth]{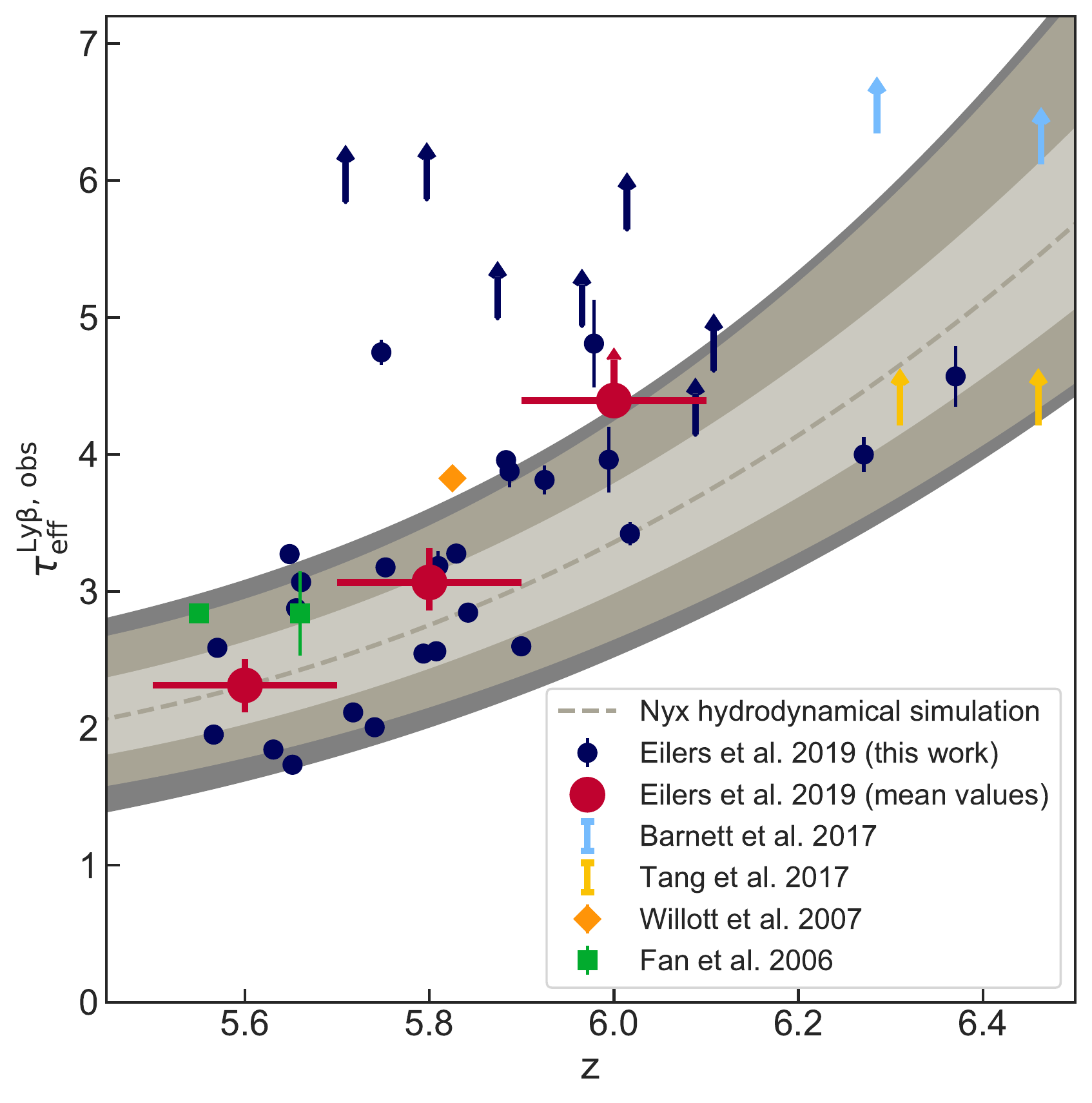}
\caption{Evolution of the observed \lyb\ optical depth with redshift. The dark blue data points show our measurements of $\tau_{\rm eff}^{\rm Ly\beta,\,obs}$, whereas the large red data points represent the opacity measurements averaged over bins of $\Delta z = 0.2$, with uncertainties determined via bootstrapping. Otherwise colored data points show all \lyb\ optical depth measurements found in the literature along quasar sightlines that are not part of our data sample. Note, however, that the chosen spectral bin size here varies between different analyses. The grey underlying region shows the predicted redshift evolution from the Nyx hydrodynamical simulation assuming a uniform UVB. We have simulation outputs in steps of $\Delta z = 0.5$ and use a cubic spline function to interpolate the shaded regions between the redshift outputs. The light and medium grey shaded regions indicate the $68$th and $95$th percentile of the scatter expected from density fluctuations in the simulations, whereas the dark grey regions show any additional scatter due to $\sim 20\%$ continuum uncertainties. \label{fig:tau}} 
\end{figure}

\subsection{The Effective Optical Depth}\label{sec:tau_eff}
We estimate the opacity of the IGM by means of the effective optical depth, i.e. 
\begin{equation}
\tau_{\rm eff} = -\ln \langle F\rangle, \label{eq:tau}
\end{equation}
where $\langle F\rangle$ denotes the continuum normalized transmitted flux averaged over discrete spectral bins along the line of sight to each quasar. In this study, we choose spectral bins of $40$~cMpc, instead of a bin size of $50$~\cmpch\ used in previous work \citep{Becker2015, Eilers2018a, Bosman2018}. This smaller bin is chosen such that the \lyb\ forest region is better sampled. 
Note that whenever the average flux $\langle F\rangle$ is negative or detected with less than $2\sigma$ significance, we adopt a lower limit on the optical depth at the $2\sigma$ level, in accordance with previous work \citep{Fan2006, Becker2015}. 

The Ly$\alpha$ forest lies between the \lya\ and \lyb\ emission lines at the rest-frame wavelengths $\lambda_{\rm Ly\alpha} = 1215.67$~{\AA} and $\lambda_{\rm Ly\beta} = 1025.72$~{\AA}, respectively, whereas the Ly$\beta$ forest covers the wavelength region between the \lyb\ and the \lyc\ (at rest-frame wavelength $\lambda_{\rm Ly\gamma} = 972.54$~{\AA}) emission lines.  
However, we do not conduct the IGM opacity measurements within the whole wavelength region, but exclude the proximity zones around each quasar as in \citet{Eilers2018a}, i.e. we mask the region around each quasar that is heavily influenced by the quasar's own radiation resulting in enhanced transmitted flux. Additionally, we choose the rest-frame wavelengths $1030$~{\AA} and $975$~{\AA} as the minimum wavelengths for our measurements in the Ly$\alpha$ and Ly$\beta$ forest, respectively. 
Thus for a typical quasar at $z_{\rm em}\sim 6$ with a proximity zone of $R_p\approx 5$~pMpc we measure the opacity of the Ly$\alpha$ and Ly$\beta$ forests between the observed wavelengths $7210-8389$~{\AA} and $6825-7078$~{\AA}, respectively. 

Additionally, as described in detail in \citet{Eilers2018a} we correct for small offsets in the zero-level of each spectrum that might have been introduced due to improper sky subtraction, and mask the spectral regions around DLAs, in order to avoid any biases in the estimation of the IGM opacity\footnote{Note that the masking of bins containing a DLA removes all bins within the Ly$\beta$ forest along two quasar sightlines, i.e. $\rm SDSS\,J0100+2802$ and $\rm SDSS\,J1148+5251$. }. 

Note that we do not attempt to correct for any metal line contamination within the \lya\ and \lyb\ forests. \citet{FaucherGiguere2008b} show in their Fig.~$8$ that the relative metal correction to $\tau_{\rm eff}$ in the \lya\ forest decreases with increasing redshift from $13\%$ at $z=2$ to $5\%$ at $z=4$. Assuming a monotonic increase in the enrichment of the IGM with time, we expect to have a negligible metal contamination at $z\approx 6$.

\section{Measurements of the IGM Opacity}\label{sec:results}

\begin{figure*}[!t]
\centering
\includegraphics[width=\textwidth]{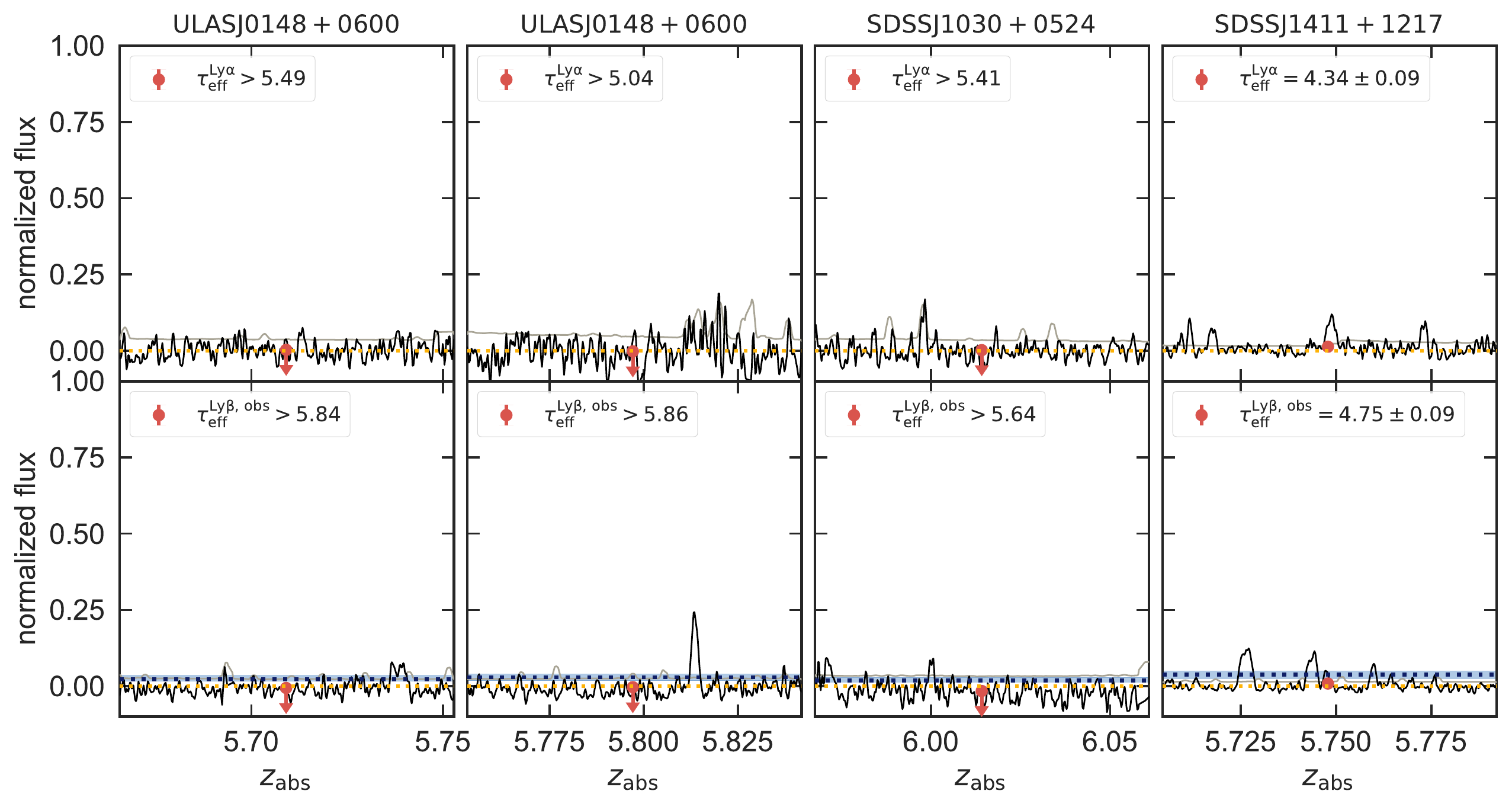}
\caption{Selected spectral bins along four different quasar sightlines showing the transmitted flux (black) and the noise vector (grey) in the \lya\ forest (top panels) and the \lyb\ forest (lower panels) in a spectral bin centered around the same absorption redshift $z_{\rm abs}$. The red data points indicate the measured mean fluxes $\langle F\rangle$ in the respective bins and the corresponding optical depths are indicated in the legend. Note that the errorbars on the mean fluxes are smaller than the symbols. The yellow dotted lines indicate the zero-level of the flux. The blue dotted line and shaded bands in the lower panels show the expected mean flux and $68$th percentile in the Ly$\beta$ forest from a reionization model with a fluctuating UVB. \label{fig:chunks}}
\end{figure*}

For a quasar at redshift $z_{\rm em}$ we encounter foreground \lya\ forest absorption from the 
IGM at lower redshift, i.e. $z_{\rm fg}$, within the observed wavelength range of the \lyb\ forest, i.e. 
\begin{equation}
z_{\rm fg} = \left((1 + z_{\rm em})\,\frac{\lambda_{\rm Ly\beta}}{\lambda_{\rm Ly\alpha}}\right) - 1.  
\end{equation}
The redshift range of the foreground \lya\ forest absorption within the \lyb\ forest is $z_{\rm fg}\approx 4.61-4.82$ for a quasar at $z_{\rm em}\approx 6$.

In this paper we calculate the effective optical depth by means of Eqn.~\ref{eq:tau} in the \lya\ forest $\tau_{\rm eff}^{\rm Ly\alpha}$ as well as the \textit{observed} effective optical depth in the \lyb\ forest $\tau_{\rm eff}^{\rm Ly\beta,\,obs}$, which includes the absorption from the foreground \lya\ forest. Hence, we do not attempt to ``correct'' the \lyb\ opacity measurements by subtracting the foreground \lya\ absorption, but rather analyze the observed \lyb\ opacity, which is the sum of the \textit{pure} \lyb\ optical depth and the \lya\ forest opacity in the foreground IGM. 
We measure both effective optical depths within the \lya\ and the \lyb\ forest in discrete spectral bins around the same central redshift $z_{\rm abs}$. Our chosen bin size of $40$~cMpc allows us to obtain two estimates along each quasar sight line, in the absence of any masked spectral bins due to intervening DLAs. 

The resulting measurements of the effective optical depths along the $19$ quasar sight lines in our data sample are listed in Tab.~\ref{tab:measurements}. 
The redshift evolution of the observed effective optical depth in the \lyb\ forest is presented in Fig.~\ref{fig:tau}. We also show the mean values, $\langle\tau_{\rm eff}^{\rm Ly\beta,\,obs}\rangle$, in bins of $\Delta z=0.2$, which are listed in Tab.~\ref{tab:comp_lyb}, as well as a compilation of all \lyb\ measurements from the literature \citep{Fan2006, Willott2007, Tang2017, Barnett2017}. We find a sharp increase in the observed \lyb\ optical depth for $z\gtrsim5.8$, which is in agreement with previous results \citep{Lidz2002, Songaila2004, Fan2006}. 

The grey bands underlying the measurements in Fig.~\ref{fig:tau} show the expected redshift evolution of the \lyb\ opacity from the \texttt{Nyx} cosmological hydrodynamical simulation (see \S~\ref{sec:sims} for details) with the default value of $\gamma=1.5$ for the slope of the \tdr \citep{Hui1997, UptonSanderbeck2016}. 
The light and medium grey bands show the expected $1\sigma$ and $2\sigma$ scatter, whereas the thin dark grey band presents any additional scatter expected from $\sim 20\%$ uncertainties in the quasar continuum estimates, which only have a small effect \citep{Becker2015, Eilers2017b}. As noticed in multiple previous studies \citep[e.g.][]{Fan2006, Becker2015, Eilers2018a, Bosman2018} the observed scatter is larger than expected from a homogeneous reionization model, in which density fluctuations alone account for the scatter in the opacities, which gave rise to models with spatial fluctuations in the UVB or the underlying temperature field, which we will introduce in \S~\ref{sec:UVB_fluct} and \S~\ref{sec:T_fluct} .

A few selected spectral bins along four different quasar sightlines, showing the transmitted flux in the \lya\ forest and the respective co-spatial region in the \lyb\ forest, are shown exemplary in Fig.~\ref{fig:chunks}. The spectral bins depicted here are chosen because they are particularly opaque in the \lyb\ forest. All other remaining spectral bins in our data sample are shown in Fig.~\ref{fig:all_chunks1} and Fig.~\ref{fig:all_chunks2} in Appendix~\ref{app:chunks}. The red data points in each panel show the \textit{measured} mean flux in the respective spectral bin, whereas the blue dotted lines and shaded regions in the bottom panels represent the \textit{expected} mean and $68$th percentile of flux in the \lyb\ forest given the corresponding $\tau_{\rm eff}^{\rm Ly\alpha}$ measurement shown in the top panels, assuming a model of the post-reionization IGM with a fluctuating UVB (see \S~\ref{sec:UVB_fluct}). The model predicts generally higher mean fluxes than measured in these selected spectral bins, which is an important issue that we will discuss in detail in \S~\ref{sec:discussion}.

\begin{deluxetable*}{lLLLLRR}
\setlength{\tabcolsep}{10pt}
\tablecaption{Mean flux measurements in the \lya\ and \lyb\ forest. \label{tab:measurements}}
\tablehead{\colhead{object} & \dcolhead{z_{\rm em}} & \dcolhead{z_{\rm start}} & \dcolhead{z_{\rm abs}} & \dcolhead{z_{\rm end}} & \dcolhead{\langle F^{\rm Ly\alpha}\rangle} & \dcolhead{\langle F^{\rm Ly\beta,\,obs}\rangle}}
\startdata
SDSSJ0002+2550 & $5.82$ & $5.699$ & $5.656$ & $5.613$ & $0.0728\pm0.0005$ & $0.0563\pm0.0005$ \\
 &  & $5.613$ & $5.570$ & $5.528$ & $0.0245\pm0.0006$ & $0.0752\pm0.0005$ \\
SDSSJ0005-0006 & $5.844$ & $5.761$ & $5.717$ & $5.674$ & $0.0582\pm0.0022$ & $0.1205\pm0.0066$ \\
 &  & $5.674$ & $5.631$ & $5.588$ & $0.0484\pm0.0027$ & $0.1580\pm0.0067$ \\
CFHQSJ0050+3445 & $6.253$ & $6.136$ & $6.088$ & $6.041$ & $0.0027\pm0.0025$ & $-0.0018\pm0.0079$ \\
 &  & $6.041$ & $5.994$ & $5.948$ & $0.0101\pm0.0031$ & $0.0190\pm0.0046$ \\
ULASJ0148+0600 & $5.98$ & $5.842$ & $5.797$ & $5.753$ & $-0.0022\pm0.0032$ & $-0.0032\pm0.0014$ \\
 &  & $5.753$ & $5.709$ & $5.666$ & $0.0028\pm0.0021$ & $-0.0059\pm0.0014$ \\
PSOJ036+03 & $6.5412$ & $6.421$ & $6.370$ & $6.320$ & $0.0010\pm0.0021$ & $0.0104\pm0.0023$ \\
 &  & $6.320$ & $6.271$ & $6.222$ & $-0.0034\pm0.0021$ & $0.0183\pm0.0024$ \\
PSOJ060+25 & $6.18$ & $6.064$ & $6.017$ & $5.971$ & $0.0623\pm0.0039$ & $0.0327\pm0.0028$ \\
 &  & $5.971$ & $5.925$ & $5.879$ & $0.0163\pm0.0081$ & $0.0221\pm0.0023$ \\
SDSSJ0836+0054 & $5.81$ & $5.695$ & $5.652$ & $5.609$ & $0.1323\pm0.0003$ & $0.1765\pm0.0006$ \\
 &  & $5.609$ & $5.566$ & $5.524$ & $0.0887\pm0.0005$ & $0.1416\pm0.0005$ \\
SDSSJ0840+5624 & $5.8441$ & $5.692$ & $5.648$ & $5.606$ & $0.0124\pm0.0013$ & $0.0379\pm0.0014$ \\
SDSSJ1030+0524 & $6.309$ & $6.156$ & $6.108$ & $6.061$ & $0.0097\pm0.0020$ & $-0.0136\pm0.0050$ \\
 &  & $6.061$ & $6.014$ & $5.968$ & $0.0018\pm0.0022$ & $-0.0156\pm0.0018$ \\
SDSSJ1137+3549 & $6.03$ & $5.874$ & $5.829$ & $5.785$ & $0.0126\pm0.0040$ & $0.0378\pm0.0021$ \\
 &  & $5.785$ & $5.741$ & $5.697$ & $0.0068\pm0.0031$ & $0.1342\pm0.0022$ \\
SDSSJ1306+0356 & $6.016$ & $5.887$ & $5.842$ & $5.797$ & $0.1467\pm0.0022$ & $0.0581\pm0.0010$ \\
 &  & $5.797$ & $5.752$ & $5.709$ & $0.0486\pm0.0015$ & $0.0418\pm0.0010$ \\
ULASJ1319+0950 & $6.133$ & $6.025$ & $5.978$ & $5.932$ & $-0.0021\pm0.0064$ & $0.0082\pm0.0026$ \\
 &  & $5.932$ & $5.887$ & $5.841$ & $0.0265\pm0.0089$ & $0.0207\pm0.0024$ \\
SDSSJ1411+1217 & $5.904$ & $5.792$ & $5.748$ & $5.704$ & $0.0131\pm0.0012$ & $0.0087\pm0.0008$ \\
 &  & $5.704$ & $5.661$ & $5.618$ & $0.0362\pm0.0011$ & $0.0465\pm0.0008$ \\
SDSSJ1602+4228 & $6.09$ & $5.929$ & $5.883$ & $5.838$ & $0.0136\pm0.0040$ & $0.0191\pm0.0013$ \\
 &  & $5.838$ & $5.794$ & $5.749$ & $0.0604\pm0.0043$ & $0.0784\pm0.0014$ \\
SDSSJ1630+4012 & $6.065$ & $5.852$ & $5.807$ & $5.763$ & $0.0366\pm0.0071$ & $0.0771\pm0.0028$ \\
SDSSJ2054-0005 & $6.0391$ & $5.946$ & $5.900$ & $5.855$ & $0.0519\pm0.0055$ & $0.0743\pm0.0047$ \\
 &  & $5.855$ & $5.810$ & $5.765$ & $0.0085\pm0.0059$ & $0.0414\pm0.0045$ \\
SDSSJ2315-0023 & $6.117$ & $6.012$ & $5.965$ & $5.919$ & $-0.0073\pm0.0040$ & $-0.0068\pm0.0036$ \\
 &  & $5.919$ & $5.874$ & $5.829$ & $-0.0135\pm0.0073$ & $-0.0038\pm0.0034$ \\
 \enddata
\tablecomments{The different columns show the name of the object and its emission redshift $z_{\rm em}$, the start of the redshift bin $z_{\rm start}$, the mean redshift of each bin $z_{\rm abs}$, and the end of each bin $z_{\rm end}$, as well as the measured mean flux of the continuum normalized spectra in the \lya\ and \lyb\ forest. }
\end{deluxetable*}

\begin{deluxetable}{LRRRc}
\vspace*{.4 cm}
\tablecaption{Measurements of the average flux and optical depth within the \lyb\ forest.  \label{tab:comp_lyb}}
\tablehead{\dcolhead{z_{\rm abs}} & \dcolhead{\langle F^{\rm Ly\beta,\,obs} \rangle} & \dcolhead{\sigma_{\langle F^{\rm Ly\beta,\,obs}\rangle}} & \dcolhead{\langle\tau_{\rm eff}^{\rm Ly\beta,\,obs}\rangle} & \dcolhead{\sigma_{\langle\tau_{\rm eff}^{\rm Ly\beta,\,obs}\rangle}}}
\startdata
5.6 & 0.0989 & 0.0193 & 2.3140 & 0.1952 \\
5.8 & 0.0466 & 0.0106 & 3.0658 & 0.2279 \\
6.0 & 0.0082 & 0.0062 & >4.3935 & --- \\
\enddata
\tablecomments{The columns show the mean redshift $z_{\rm abs}$ of the redshift bins of size $\Delta z = 0.2$, the averaged flux $\langle F^{\rm Ly\beta,\,obs}\rangle$ and its uncertainty $\sigma_{\langle F^{\rm Ly\beta,\,obs}\rangle}$ determined via bootstrapping, and the mean optical depth $\langle\tau_{\rm eff}^{\rm Ly\beta,\,obs}\rangle$ in that redshift bin and its error $\sigma_{\langle\tau_{\rm eff}^{\rm Ly\beta,\,obs}\rangle}$, also determined via boostrapping. }\end{deluxetable}

\section{Models of the Post-Reionization IGM}\label{sec:sims}

Several previous studies have shown that the scatter in the \lya\ optical depth measurements of the IGM cannot be explained by fluctuations in the underlying density field alone, but rather it requires a spatially inhomogeneous reionization scenario with additional large-scale fluctuations in the UVB or the temperature field to reproduce the observations \citep{Becker2015, Daloisio2018_Tfluct, Davies2018a, Eilers2018a, Bosman2018}. 
In this work we compare these different physical models of the post-reionization IGM to the co-spatial opacity measurements in the \lya, as well as the \lyb\ forest. 

All reionization models that we consider make use of the \texttt{Nyx} cosmological hydrodynamical simulation with 100 \cmpch\ on a side and 4096$^3$ dark
matter particles and gas elements on a uniform Eulerian grid, which was designed for precision studies of the \lya\ forest \citep{Almgren2013, Lukic2015}. We extract skewers of density,
temperature, and velocity along the directions of the grid axes from
simulation outputs at $3.0 \leq z\leq 6.5$ in steps of $\Delta z =
0.5$.  The outputs at lower redshifts are used to model the
contamination of the \lyb\ forest opacity by foreground
\lya\ absorption, which we will describe in \S~\ref{sec:lya_tau}.  For
redshifts in between the simulation outputs, we take the closest
output and re-scale the density field by $(1+z)^3$ accordingly. The
simulation adopts the uniform UVB model from \citet{HaardtMadau2012},
resulting in an IGM model which (uniformly) reionized at early times,
i.e. $z_{\rm reion}>10$ \citep{Lukic2015, Onorbe2017a}. 
In Appendix~\ref{app:resolution} we perform a set of convergence tests for the optical depth in the \lyb\ forest at high redshift. The convergence of the \lya\ forest has been tested previously in \citet{Onorbe2017}. 

First, we model the post-reionization IGM with a spatially inhomogeneous UVB
(\S~\ref{sec:UVB_fluct}), and second, we assume a spatially fluctuating
temperature field in the IGM (\S~\ref{sec:T_fluct}). We will see that neither of these models reproduces our observations very well, and also consider a model with a uniform UVB, but vary the slope of the temperature-density relation of the IGM (\S~\ref{sec:nyx}). 
For each model we will then calculate the optical depths along skewers through the simulation
box (\S~\ref{sec:lya_tau}) and forward-model the spectral noise of
our data set onto the skewers (\S~\ref{sec:forward-model}).

\subsection{Fluctuating UVB}\label{sec:UVB_fluct}

We first compare our measurements to predictions of the \lyb\ forest
opacity from a published reionization model with a spatially inhomogeneous UVB by
\citet{Davies2018a}. The model consists of UVB fluctuations from an independent ``semi-numerical"
simulation following \citet{DaviesFurlanetto2016}, which we summarize briefly here.
First, using the \texttt{21cmFAST} code, we generated a set of cosmological initial conditions in a ($400$~cMpc)$^3$ volume. From these initial conditions we derived the locations of dark matter halos following the excursion set method of \citet{MesingerFurlanetto2007}, and shifted their positions to $z=6$ with the Zel'dovich approximation \citep{Zeldovich1970}. The UV luminosities of the halos were then computed via abundance matching (e.g. \citealt{ValeOstriker2004}) to the \citet{Bouwens2015} UV luminosity function. Finally, the ionizing radiation field was computed assuming a halo mass-independent conversion from UV luminosity to ionizing luminosity, and a spatially-varying mean free path of ionizing photons \citep{DaviesFurlanetto2016}. We then impose the fluctuating UVB onto the \texttt{Nyx} skewers and re-compute the ionization state of the gas assuming ionization equilibrium. 

Because the UVB fluctuations were computed in an independent cosmological volume from the hydrodynamical simulation, this approach modestly overestimates the effect of UVB fluctuations on the IGM opacity because the anti-correlation between the UVB and the large-scale density field \citep{DaviesFurlanetto2016, Davies2018b} is lost.  While small-scale correlations between the
UVB and the density field may alter the relationship between \lya\ and
\lyb\ forest opacity \citep{OhFurlanetto2005}, the UVB fluctuations in
the \citet{Davies2018b} model manifest on scales comparable to the
size of our $40$ cMpc bins, so we do not expect a substantial effect.

\subsection{Fluctuating Temperature Field of the IGM}\label{sec:T_fluct}

In order to model temperature fluctuations we use a new hybrid method introduced in \citet{Onorbe2018} that uses a small set of phenomenological input parameters, and combines a semi-numerical reionization model \citep[\texttt{21cmFAST};][]{Mesinger2011} to solve for the topology of reionization, as well as an approximate model of how reionization heats the IGM, with the cosmological hydrodynamical code \texttt{Nyx}. 
Instead of applying a uniform UVB to the whole simulation box, which was the procedure in the original \texttt{Nyx} simulations, the UVB is now applied to denser regions in the simulations at earlier reionization redshifts $z_{\rm reion}$, whereas the underdense voids will be reionized at later times. At the same time, the gas temperature is heated up. The topology of the different reionization times $z_{\rm reion}$ are extracted from the semi-numerical model. 
 
This guarantees that the temperature evolution of the inhomogeneous reionization process and the small-scale structure of the diffuse gas of the IGM is resolved  and captured self-consistently. In this work we used the IR-A reionization model presented in \citet{Onorbe2018}, which is an extended reionization model consistent with the  observations of the cosmic microwave background (CMB) by \citet{Planck2018} with a median reionization redshift of $z_{\rm reion} = 7.75$, a duration of the reionization process of $\Delta z_{\rm reion}=4.82$, and a heat injection due to reionization of $\Delta T=20,000$~K. We apply this model to a \texttt{Nyx} simulation with L=$40$~Mpc$\,h^{-1}$ box side and $2048^3$ resolution elements. The full thermal evolution of this model at the redshifts of interest for this work can be seen in Fig. $5$ and $6$ in \citet{Onorbe2018}.

\subsection{Varying the Slope of the Temperature-Density Relation in a Uniform IGM}\label{sec:nyx}

We will see in \S~\ref{sec:discussion} that neither of the two models described above
provides a good fit to our opacity measurements in \textit{both} the \lya\ and
the \lyb\ forests. Since the ratio of these opacities is sensitive to the temperature-density relation of the IGM, we will also compare our measurements to models with a homogeneous IGM, but different slopes $\gamma-1$ of the temperature-density relation (see Eqn.~\ref{eq:tdr}). 

To this end, we simply ignore the temperatures of the extracted \texttt{Nyx} skewers and calculate new temperatures in post-processing for the gas densities along each skewer by means of Eqn.~\ref{eq:tdr}, assuming $T_0=10,000$~K. We then take the velocity skewers and map again the modified real space into redshift space to calculate mean fluxes and opacities.

\subsection{Calculating the Optical Depth from Simulated Skewers}\label{sec:lya_tau}

We will compare our observations to the reionization models in three redshift bins with
$\Delta z = 0.2$ at the central redshifts of $z=5.6$, $z=5.8$, and
$z=6.0$. At each redshift we extract $N_{\rm skew} = 2000$ skewers from the simulation box of the same size as our measurements, i.e. $40$~cMpc.
It is well known that the Ly$\alpha$ opacity depends on the unknown amplitude of the UVB $\Gamma_{\rm HI}$, thus we re-scale the optical depth in the \lya\ forest to match the observations and determine this unknown parameter.
For each skewer we compute the mean flux, or equivalently the effective optical depth according to
\begin{align}
\langle F^{\rm Ly\alpha}\rangle &= \langle\exp\left[-\tau_{i}^{\rm Ly\alpha}\right]\rangle\nonumber\\
& = \langle \exp\left[-A_0 \times \tau_{i}^{\rm Ly\alpha, unscaled}\right]\rangle, \label{eq:rescale}
\end{align}
where the angle brackets denote the average in the $40$~cMpc bins, and in the last equality we introduce the scaling factor $A_0$ which
provides the lever arm for tuning the photoionization rate of the UVB $\Gamma_{\rm HI}$ to match the observations. 

Studies of the Ly$\alpha$ forest at lower redshift, i.e. $2<z<5$,
typically tune $A_0$ to match the mean flux. Since this study focuses
on comparing the distribution of mean fluxes (effective optical
depths) in $40$~cMpc bins, we instead determine the value of $A_0$ by
requiring the $25$th percentile of the cumulative distribution of the mean flux to
match the data. Specifically, for the data in any redshift bin we can
solve the equation $P( > \langle F^{\rm Ly\alpha}\rangle) = 0.25$ for
$\langle F^{\rm Ly\alpha}\rangle$.  Our procedure for
setting the unknown $\Gamma_{\rm HI}$ for any given model is then to
vary $A_0$ until the model matches the $\langle F^{\rm
  Ly\alpha}\rangle$ determined from the data for that
redshift bin. From the cumulative distribution of values in
Tab.~\ref{tab:measurements_lya} we measure $\langle F^{\rm Ly\alpha}\rangle_{\rm
  25th}(z=5.6)\approx 0.0713$, $\langle F^{\rm Ly\alpha}\rangle_{\rm
  25th}(z=5.8)\approx 0.0363$, and $\langle F^{\rm
  Ly\alpha}\rangle_{\rm 25th}(z=6.0)\approx 0.0144$.  Note
that the choice of the $25$th percentile is somewhat arbitrary, and we
could have also chosen to match the observed mean or median flux, which would not change any of the conclusions of this work.

We then obtain the \textit{pure} \lyb\ optical depth $\tau_{i}^{\rm
  Ly\beta}$ at each pixel from $\tau_{i}^{\rm Ly\alpha}$ by scaling
according to the oscillator strengths of their transitions and the
respective wavelengths, i.e.
\begin{equation}
\tau_{i}^{\rm Ly\beta} = \frac{f_{\rm Ly\beta}}{f_{\rm Ly\alpha}}\,\frac{\lambda_{\rm Ly\beta}}{\lambda_{\rm Ly\alpha}}\,\tau_{i}^{\rm Ly\alpha}\label{eq:oscillator}
\end{equation}
with $f_{\rm Ly\alpha} = 0.41641$ and $f_{\rm Ly\beta} = 0.079142$ \citep[Table $4$ in][]{WieseFuhr2009}. This scaling is correct provided that there are no damping wing effects due to the presence of very dense or highly neutral gas.
Given that our analysis is focused on the low density IGM at $z\gtrsim 5.5$ probed by the \lya\ and \lyb\ forests at typical locations of the universe, this is a good approximation.

However, the \textit{observed} \lyb\ effective optical depth $\tau_{\rm eff}^{\rm Ly\beta,\,obs}$ will be higher than the pure \lyb\ optical depth due the additional absorption from the \lya\ forest at the foreground redshift. Thus we create additional skewers along \textit{different} lines of sight with \lya\ forest absorption at the foreground redshift $z_{\rm fg}$, i.e. $\tau_{i}^{\rm Ly\alpha, fg}$,
using lower redshift outputs from the \texttt{Nyx} simulation. These lower redshift skewers are also re-scaled but to match the mean flux to be consistent with published measurements. We use the fitting formula presented in
\citet{Onorbe2017a} to the mean flux measurements by \citet{FaucherGiguere2008b}, i.e. $\langle F^{\rm Ly\alpha} \rangle (z_{\rm fg} = 4.57) \approx 0.2371$, $\langle F^{\rm Ly\alpha} \rangle (z_{\rm fg} = 4.74) \approx 0.1945$, and $\langle F^{\rm Ly\alpha} \rangle (z_{\rm fg} = 4.91) \approx 0.1560$ for the redshift bins at $z=5.6$, $z=5.8$, and $z=6.0$, respectively.

For each skewer of the high redshift \lyb\ forest, we now draw a \textit{random} \lya\ forest segment from a simulation output at the corresponding foreground redshift. In this way we ensure that we account for the full distribution of foreground \lya\ opacities, and do not underestimate the scatter between different foreground sightlines. 
At each pixel we then sum the pure \lyb\ forest optical depth with the foreground \lya\ optical depth to obtain the flux one would observe in the \lyb\ forest, i.e.
\begin{align}
  F_{i}^{\rm Ly\beta, obs} = \exp\left[-(\tau_{i}^{\rm Ly\beta} +\tau_{i}^{\rm Ly\alpha,\,fg})\right]. 
\end{align}
We calculate the observed effective \lyb\ optical depth by averaging the flux of all pixels in the spectral bin, i.e. $\tau_{\rm eff}^{\rm Ly\beta,\,obs} = -\ln\,\langle F_{i}^{\rm Ly\beta, obs}\rangle$. 

\begin{figure}[!t]
\centering
\includegraphics[width=.47\textwidth]{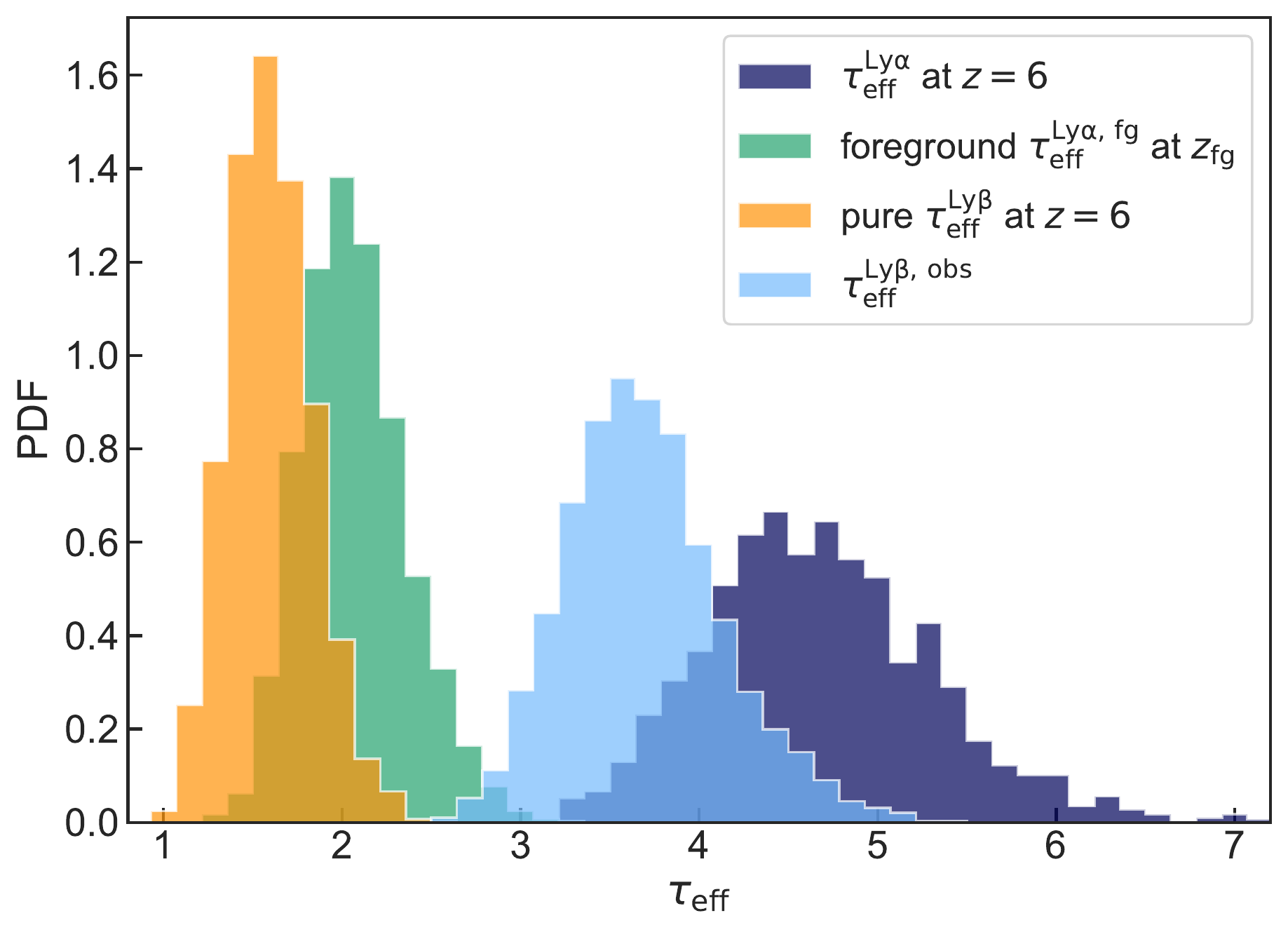}
\caption{Distribution of $\tau_{\rm eff}$ values from $2000$ skewers of the noise-free uniform reionization model with $\gamma = 1.5$ at $z=6$. The histograms show the distribution of $\tau_{\rm eff}^{\rm Ly\alpha}$ at $z=6$ (dark blue), the distribution of $\tau_{\rm eff}^{\rm Ly\alpha,\, fg}$ at the foreground redshift (green), the pure $\tau_{\rm eff}^{\rm Ly\beta}$ values (yellow), and finally the distribution of observed $\tau_{\rm eff}^{\rm Ly\beta,\,obs}$ values, after accounting for the distribution of foreground \lya opacities (light blue). \label{fig:distribution_lyb}}
\end{figure}

Fig.~\ref{fig:distribution_lyb} shows the various distributions of effective optical depths in an example case for the uniform reionization model with $\gamma=1.5$ at $z=6$, however, all other models look qualitatively similar. The dark blue histogram shows the distribution of $\tau_{\rm eff}^{\rm Ly\alpha}$ at $z=6$, whereas the green histogram depicts the distribution of $\tau_{\rm eff}^{\rm Ly\alpha,\, fg}$ along different skewers at the foreground redshift $z_{\rm fg}\approx 4.9$. The yellow histogram shows the pure $\tau_{\rm eff}^{\rm Ly\beta}$ values obtained from re-scaling accoring to Eqn.~\ref{eq:rescale}, while the light blue histogram represents the distribution of observed $\tau_{\rm eff}^{\rm Ly\beta\,obs}$ values, after accounting for the distribution of foreground \lya opacities.  

\begin{figure*}[t!]
\centering
\includegraphics[width=.95\textwidth]{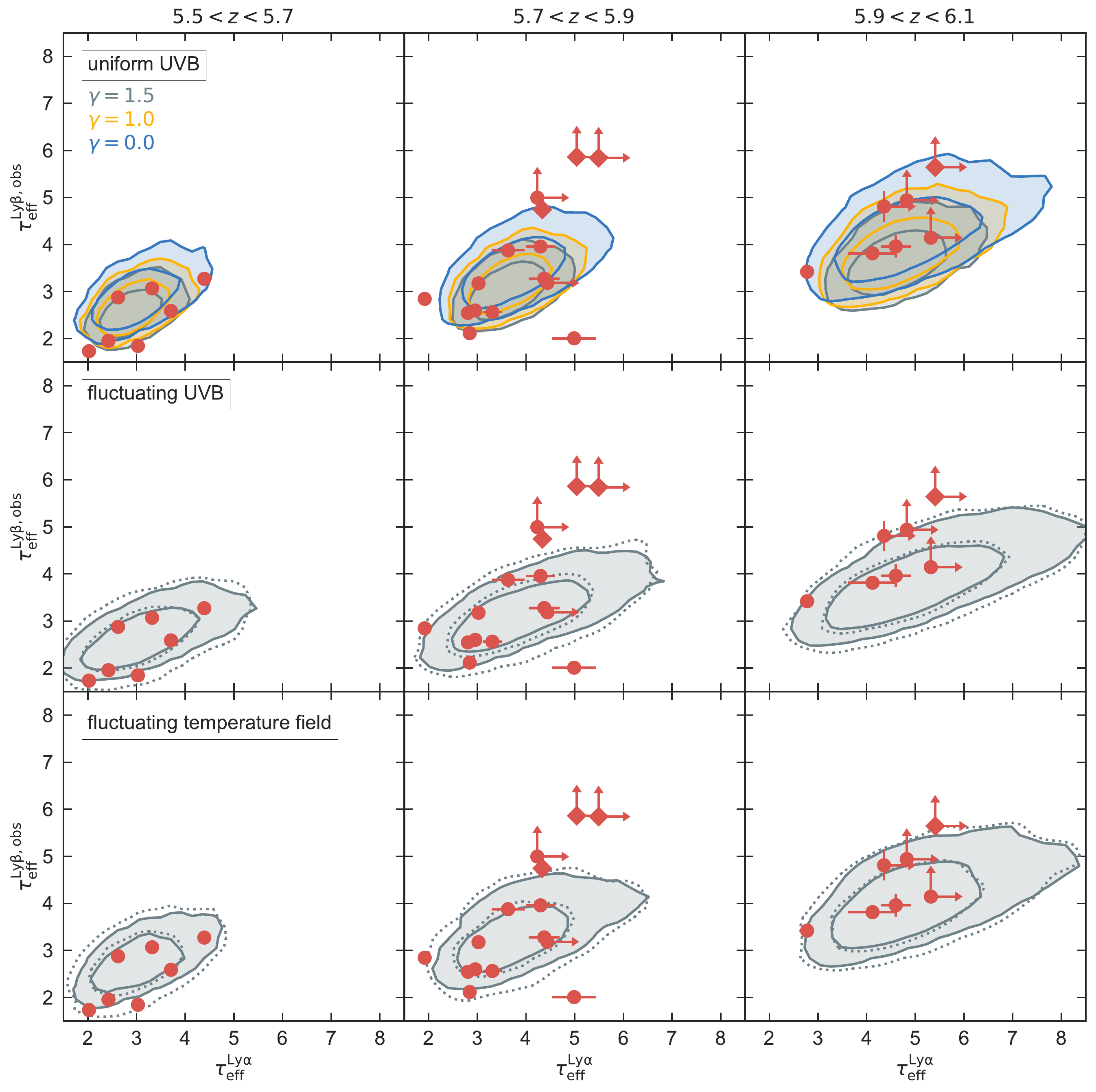}
\caption{Comparison of our \lya\ and \lyb\ opacity measurements shown as the red data points in different redshift bins, i.e. $5.5<z<5.7$ (left), $5.7<z<5.9$ (middle), and $5.9<z<6.1$ (right), to predictions from the Nyx hydrodynamical simulation post-processed in several different ways.  The contours in the top panels show the prediction from simulations with uniform UVB and different slopes of the \tdr of the IGM, whereas the middle and bottom panels show predictions from models with a fluctuating UVB or a fluctuating temperature field, respectively. Inner and outer contours show the $68$th and $95$th percentile of the distribution. The dotted contours show the respective distributions including $\sim20\%$ continuum uncertainties (which we omitted in the top panels for better readability). The data points marked as diamonds correspond to the spectral bins shown in Fig.~\ref{fig:chunks}. \label{fig:sims}}
\end{figure*}

\begin{figure*}[t!]
\centering
\includegraphics[width=\textwidth]{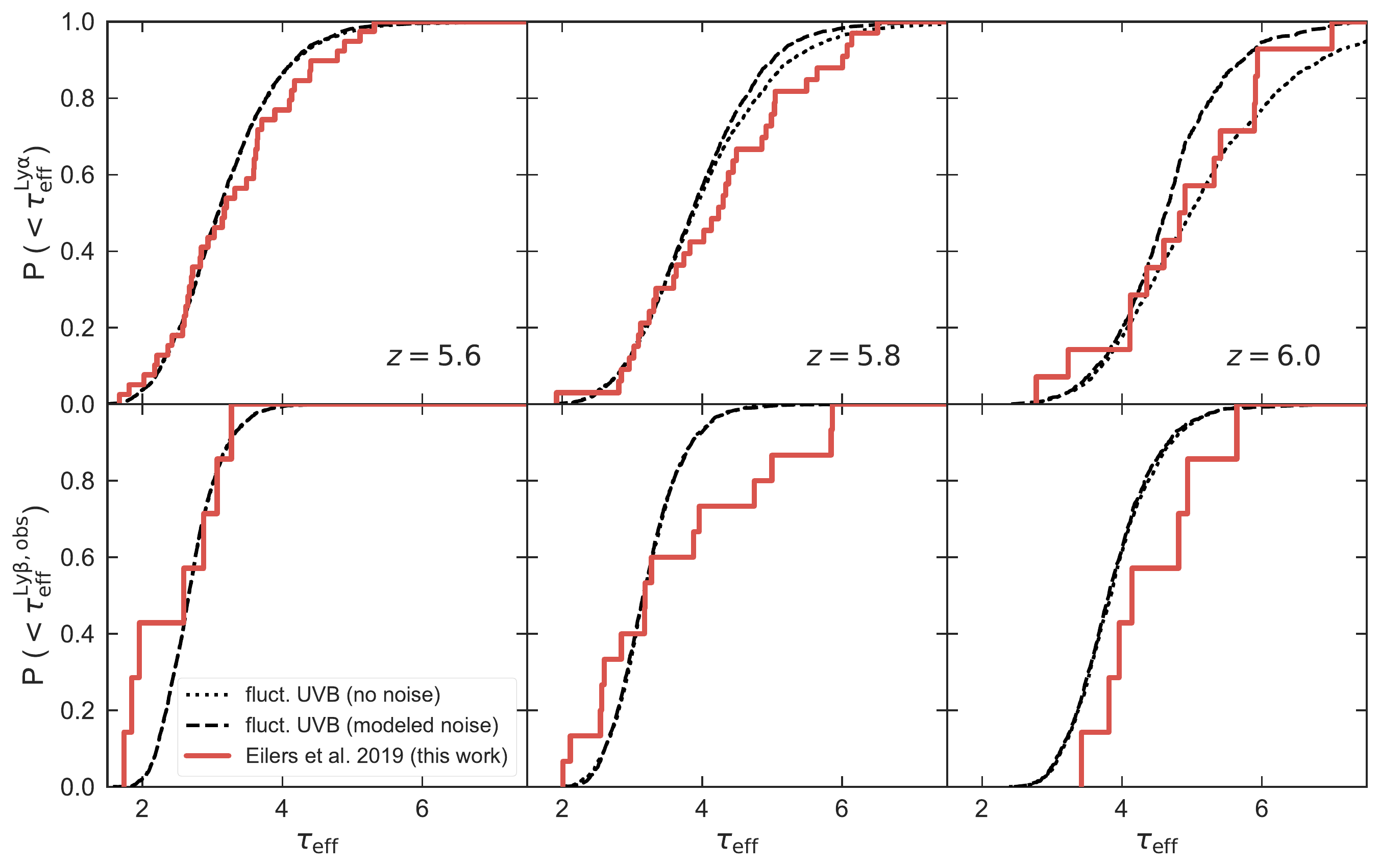}
\caption{CDFs of $\tau_{\rm eff}^{\rm Ly\alpha}$ (top panels) and $\tau_{\rm eff}^{\rm Ly\beta,\,obs}$ (lower panels) in the same redshift bins as in Fig.~\ref{fig:sims}. Our measurements are shown in red, the noise-free predictions from the reionization model with a fluctuating UVB is shown as the black dotted line, whereas the black dashed curves show the same model, now including forward-modeled spectral noise. The models are scaled to match the $25$th percentile of the observed $\tau_{\rm eff}^{\rm Ly\alpha}$ distribution. 
Note that the top panels do not only contain the measurements of $\tau_{\rm eff}^{\rm Ly\alpha}$ that have a corresponding $\tau_{\rm eff}^{\rm Ly\beta,\,obs}$ measurement at the same redshift, but rather all measurements within the \lya\ forest along all $19$ quasar sight lines. 
\label{fig:cdf}} 
\end{figure*}

\begin{figure*}[t!]
\centering
\includegraphics[width=\textwidth]{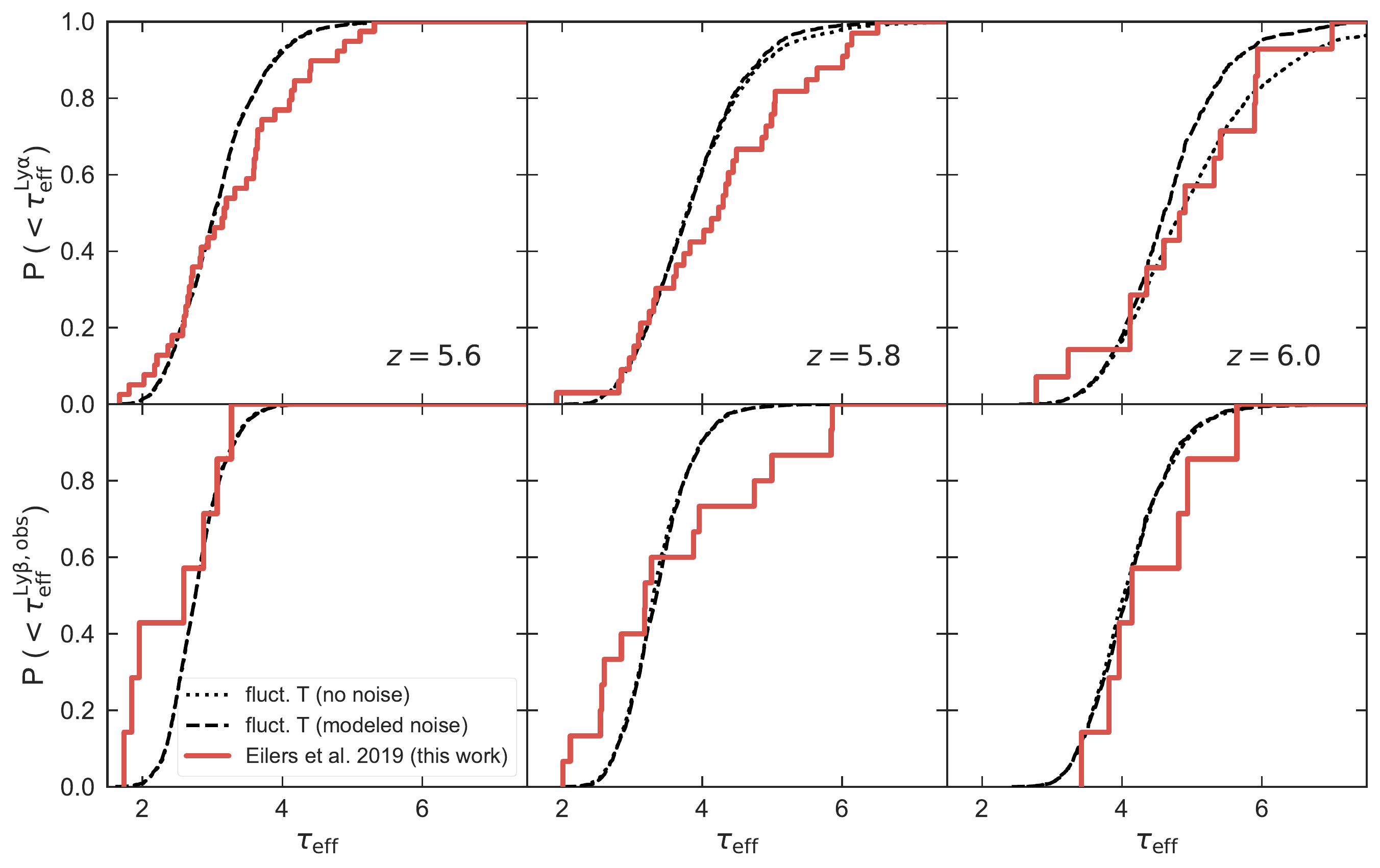}
\caption{Same as Fig.~\ref{fig:cdf} for the reionization model with a fluctuating temperature field. 
\label{fig:cdf_fluctT}} 
\end{figure*}

\begin{figure*}[t!]
\centering
\includegraphics[width=\textwidth]{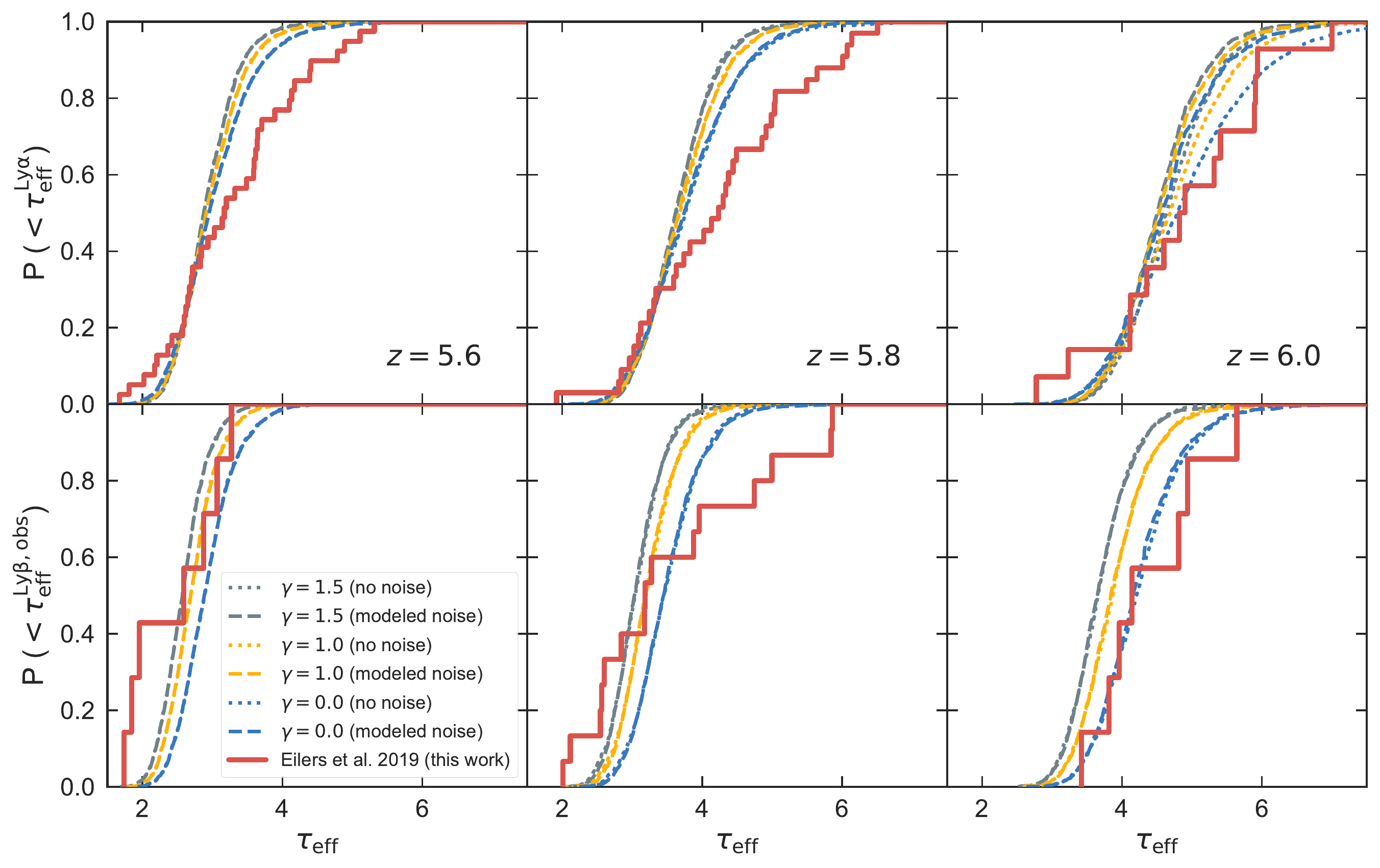}
\caption{Same as Fig.~\ref{fig:cdf} for the reionization model with a uniform UVB but different slopes of the temperature-density relation of the IGM. 
\label{fig:cdf_uni}} 
\end{figure*}

\subsection{Forward-Modeling of Spectral Noise}\label{sec:forward-model}

The spectral noise in the data influences the opacity
measurements. This is because whenever the average flux is detected
with less than $2\sigma$ significance, we adopt a lower limit on the
effective optical depth measurement (see \S~\ref{sec:tau_eff}). To mimic the
effects of the noisy data, for each mean flux value $\langle F_{\rm
  sim}\rangle$ from the simulated skewers we randomly draw an
uncertainty $\sigma_{\langle F_{\rm data}\rangle}$ from the
measurements from our data set. 
Note that formally the noise level depends on the signal, but in the limit where the objects are fainter than the sky, which is almost always the case for the highly absorbed forests in high redshift quasars, the noise is dominated by the sky background and read noise rather than the object photon noise, so this is a good approximation. 

For this mock data with forward-modeled noise we then apply the same criterion as for the real data and adopt a measurement only if $\langle F_{\rm sim}\rangle \geq 2\sigma_{\langle F_{\rm data}\rangle}$, or a lower limit at $2\sigma_{\langle F_{\rm data}\rangle}$ otherwise. This reduces the sensitivity to very high opacities in the simulated skewers, which we do not have in noisy data.

\section{Discussion}\label{sec:discussion}

\subsection{Comparison to Inhomogeneous Reionization Models}

We compare our co-spatial measurements of $\tau_{\rm eff}^{\rm Ly\alpha}$ and 
$\tau_{\rm eff}^{\rm Ly\beta,\,obs}$ to predictions from the various
models of the post-reionization IGM in three different redshift bins centered around $z=5.6,\,5.8$ and $6.0$ in Fig.~\ref{fig:sims}. The contours in the middle and bottom panels show the predicted parameter space from models with spatial fluctuations in the UVB and the temperature field, 
respectively. 
Note that we did not yet forward-model the spectral noise in the data to the
simulations here, because we are overlaying the data with error bars and indicating limits, thus, for this qualitative comparison, forward-modeling the noise would amount to effectively double counting the noise. 

While our measurements fall within the predicted parameter space in the lowest redshift bin, it is evident that the observed $\tau_{\rm eff}$ measurements in the two higher redshift bins are not well confined to the predicted parameter space. The measured $\tau_{\rm eff}^{\rm Ly\beta,\,obs}$ values show a larger scatter than expected by the models at $z\sim 5.8$, as well as an offset in the mean at $z\sim 6$.

Another way of illustrating this comparison is by means of the cumulative distribution function (CDF). In Fig.~\ref{fig:cdf} and \ref{fig:cdf_fluctT} we present the CDFs of our opacity measurements as the red curves, as well as the predicted
distributions from the post-reionization models including a
fluctuating UVB and a fluctuating temperature field, respectively, as dotted black curves. Note that whereas the measurements presented in Fig.~\ref{fig:sims} are restricted to the spectral regions where we have co-spatial measurements of $\tau_{\rm eff}^{\rm Ly\alpha}$ and $\tau_{\rm eff}^{\rm Ly\beta,\,obs}$, the cumulative histograms for $\tau_{\rm eff}^{\rm Ly\alpha}$ in Fig.~\ref{fig:cdf} and  \ref{fig:cdf_fluctT} show \textit{all} measurements from the full \lya\ forests along all quasar sightlines in our data set\footnote{Whereas in \citet{Eilers2018a} we used $50$~\cmpch bins, we recomputed the \lya\ optical depths in $40$~cMpc for the comparison here. These results are presented in Tab.~\ref{tab:measurements_lya} in Appendix~\ref{app:chunks}. }.

Both CDFs show clearly that when fine-tuning the reionization models to match the observed transmission in the \lya\ forest, the effective optical depth in the \lyb\ forest is underestimated in the highest redshift bin at $z\sim 6$. Additionally, as noted before, we find an large increase in the scatter of the measurements, most notably at $z\sim 5.8$, compared to the predictions from the models. 

Note that in the two highest redshift bins in Fig.~\ref{fig:cdf}, the difference between the observed mean Ly$\beta$ forest optical depth and the reionization model with a
fluctuating UVB is $\Delta\tau^{\rm Ly\beta}_{\rm eff}\gtrsim 1$,
which corresponds to a factor of $\gtrsim 2.5$ in the mean
flux. Hence, systematic uncertainties in the data that might arise from
poor quasar continuum estimates or issues in the data reduction
process (see also Appendix~\ref{app:photometry}), would have to change
the observed averaged flux by $\gtrsim 250\%$ to account for this
offset. This seems highly unlikely, since there is no obvious reason
why such errors, if they were present, would be so asymmetric causing the flux in the \lyb\ region to be systematically lower by this
large factor. Furthermore, if such systematics were present they would
presumably also impact the \lya\ measurements. However, a comparison
of independent measurements of the distribution of $\tau_{\rm eff}^{\rm
  Ly\alpha}$ between \citet{Bosman2018} and \citet{Eilers2018a} \citep[see Fig.~$7$ in][]{Eilers2018a} shows no evidence for systematic offsets of $\Delta\tau^{\rm Ly\beta}_{\rm eff}\gtrsim 1$. 
Note that the difference in Fig.~\ref{fig:cdf_fluctT} between the mean
\lyb\ optical depth of the fluctuating temperature model predictions
and the measurements is smaller, but still significant. \\

\subsection{Evidence for an Inverted Temperature-Density Relation of the IGM}

We know that the ratio of \lya\ and \lyb\ optical depths is sensitive to the ionization as well as the thermal state of the IGM \citep{OhFurlanetto2005, Trac2008, FurlanettoOh2009}. Thus by varying the slope of the \tdr of the intergalactic gas and adjusting it to be more isothermal (or inverted), we expect an increase in the predicted $\tau_{\rm eff}^{\rm Ly\beta,\,obs}$. 

To this end, we also compare our observations to reionization models with a uniform IGM but different slopes of their temperature-density relations, i.e. the fiducial value of $\gamma = 1.5$, an isothermal \tdr with $\gamma=1.0$, and a highly inverted slope with $\gamma = 0.0$. 
This is shown in the top panels of Fig~\ref{fig:sims}, as well as in Fig.~\ref{fig:cdf_uni}. 

The comparison shows that the model with a highly inverted temperature-density relation, i.e. $\gamma = 0.0$, predicts a higher \lyb\ opacity and seems to match the observations in the \lyb\ forest in the highest redshift bin better than the previously discussed inhomogeneous reionization models, although it is still not fully capturing the whole parameter space of the observations. 
Additionally, all uniform reionization models do not show enough spatial fluctuations to predict the  scatter in the \lya\ opacity. 

How likely is it that the IGM follows an inverted \tdr at $z\sim 6$? 
It has been argued that reionization events dramatically alter the
thermal structure of the IGM, increasing the IGM temperature by at
least an order of magnitude up to $T=25,000-30,000$~K
\citep{FurlanettoOh2009, Daloisio2018_Tfluct}. Several hydrodynamical simulations that model the reionization process self-consistently have shown that a flat or inverted \tdr arises naturally during reionization events due to the different number densities of ionizing sources in over-- and underdense regions \citep[e.g.][]{Trac2008, Finlator2018}, i.e. the gas in dense regions ionizes first due to the higher number of ionizing sources and have thus more time to cool down, whereas low-density voids are ionized last and hence contain the hottest gas at the end of reionization, which 
leads to an at least partially inverted temperature-density relation of the IGM 
\citep{FurlanettoOh2009, LidzMalloy2014, Daloisio2015,
  Daloisio2018_Tfluct}. \citet{Onorbe2018} modeled this process in detail
and show that their model (see \S~\ref{sec:T_fluct}) has an average slope of the
temperature-density relation of $\gamma \approx 1.1$, i.e. roughly
isothermal, at $z\sim 6$, but with a very large scatter (see their
Fig.~$5$), which arises due to a superposition of different heat
injections and subsequent cooling at different times. Thus their model
predicts that at least some regions of the IGM do indeed follow an inverted temperature-density relation. This could explain why the model with a fluctuating temperature field (Fig.~\ref{fig:cdf_fluctT}) matches the $\tau_{\rm eff}^{\rm Ly\beta,\,obs}$ observations better than the model with fluctuations in the UVB (Fig.~\ref{fig:cdf}). 

Although \citet{FurlanettoOh2009} claim that rapid adiabatic and Compton cooling will quickly erase any inversion of the thermal state of the IGM after the reionization event, we might still be observing the end stages of this process at $z\sim 5.5-6.0$. This result complies with the conclusions from \citet{Trac2008} who, contrary to \citet{FurlanettoOh2009}, argue that the inversion of the IGM endures long past the reionization process \citep[see also][]{Finlator2018}. 
Thus, our results provide support for a late, and extended EoR, which is also in good agreement with several recent studies \citep{Eilers2018a, Bosman2018, Planck2018, Kulkarni2019}.

\subsection{Evidence for Stronger Fluctuations in the Opacity of the \lya\ Forest}

In previous studies comparisons of $\tau_{\rm eff}$ measurements to reionization models have been performed assuming an infinite S/N ratio of the simulated skewers. These noise-free predictions for the CDFs of $\tau_{\rm eff}$ correspond to the dotted
curves in Fig.~\ref{fig:cdf}, \ref{fig:cdf_fluctT}, and
\ref{fig:cdf_uni}. However, in order to conduct a proper comparison between the noisy data of any realistic data set to these models, it is 
required that one models the spectral noise. Thus, as discussed in
\S~\ref{sec:forward-model}, we forward-model our observations by
adding spectral noise to the simulated skewers. 

The models including the forward-modeled noise are shown as the dashed curves in Fig.~\ref{fig:cdf}, \ref{fig:cdf_fluctT}, and \ref{fig:cdf_uni}, which predict less scatter at high optical depths. As a result the models with a spatially inhomogeneous reionization process that had been claimed previously to fully reproduce the scatter in the \lya\ opacity observations (Fig.~\ref{fig:cdf} and \ref{fig:cdf_fluctT}) do \textit{not} contain enough spatial fluctuations once we account for the spectral noise in the data. This discrepancy is most obvious in the top right panel of each figure for $\tau_{\rm eff}^{\rm Ly\alpha}$ measurements at $z=6.0$. 

The reason for this discrepancy is quite simple: In any realistic dataset there will be a distribution in the S/N ratios of the spectra due to the various object magnitudes and exposure times, and this noise manifests as noisy $\tau_{\rm eff}$ measurements. In order to detect a very high $\tau_{\rm eff}$, one
must not only encounter a rare fluctuation in the IGM along a quasar sightline, but
also this fluctuation must be probed by a sufficiently bright quasar or deep enough spectrum to
measure a high $\tau_{\rm eff}$. Thus large $\tau_{\rm eff}$ fluctuations in the IGM need to be more common than one naively expects from infinite S/N models. 

We conclude that the forward-modeling of spectral noise is essential to conduct a fair data-model comparison, and our results suggest that stronger spatial inhomogeneities than previously assumed are needed to fully capture the observations. At lower opacities the difference between perfect and noisy models is less significant, since most measurements are actual high S/N ratio detections rather than lower limits. \\\\

\section{Summary}\label{sec:summary}

In this paper we measured the IGM opacity in the \lya\ as well as the \lyb\ forest along $19$ quasar
sightlines. Due to the different oscillator strengths of their transitions, the \lya\ and \lyb\ optical depth depend on different gas densities, temperatures, and photoionization rates. Thus the relation between \lya\ and \lyb\ optical depths provides new stringent constraints on the ionization and thermal state of the post-reionization IGM. A comparison of our measurements to several reionization models that we obtain by post-processing the state-of-the-art \texttt{Nyx} hydrodynamical simulation, reveals two main results: 

\begin{itemize}
\item  It is essential to account for the spectral noise in the data to conduct a fair comparison between the simulations and the observations. When forward-modeling the spectral noise to the simulated skewers, the sensitivity to high opacity regions becomes reduced. This arises from the fact that in order to measure a high opacity region in the universe along a quasar sightline, a high quality spectrum of this quasar is required with a high $\rm S/N$ ratio, since otherwise a lower limit on the mean flux is adopted. Thus in a realistic data set with different $\rm S/N$ ratios due to different exposure times and quasar luminosities, the sensitivity to high opacity regions is lower than in models with infinite $\rm S/N$ skewers. Hence forward-modeling the noise to the simulations effectively decreases the scatter in optical depths predicted by the simulations. Thus, inhomogeneous reionization models including spatial UVB fluctuations or a fluctuating temperature field, which have been believed to fully reproduce the $\tau_{\rm eff}^{\rm Ly\alpha}$ measurements in previous studies, do not predict enough scatter between different quasar sightlines to match our observations. 
Hence, our results provide evidence for stronger spatial fluctuations in the reionization models to fully capture the observations than all previous studies assumed. 

\item All current reionization models underpredict the opacity in the \lyb\ forest, when fine-tuning them to match the observations in the \lya\ forest. The difference increases with increasing redshift, when approaching the EoR. 
Models with a fluctuating temperature field seem to match the \lyb\ optical depths better than models with a fluctuations in the UVB, but it may require an even more inverted \tdr of the IGM to fully capture the observed ratio of \lya\ and \lyb\ opacities at $z\sim 6$. 
\end{itemize}

Whether or not the stronger spatial fluctuations, required to match the observed opacities in the \lya\ forest, would produce models which also match the yet underpredicted \lyb\ optical depths, remains to be seen. If temperature fluctuations dominate and are stronger than current simulations suggest, then this could result in an IGM with an inverted \tdr that might make the models more consistent with our measurements. Future models that encompass effects of fluctuations in the UVB as well as the temperature field \citep[see discussion in][]{Onorbe2018} could account for the fact that the reionization process ended relatively late \citep[e.g.][]{Kulkarni2019}. It is not yet obvious how the interplay between these various effects will change the predictions for the \lya\ as well as the \lyb\ optical depths. Thus, further modeling will be required to explain the observed opacity both in the \lya\ as well as the \lyb\ forest, which will hopefully lead to new insights regarding the morphology, timing, and thermal evolution of the reionization process.

\acknowledgments

The authors would like to thank the referee for their very helpful and constructive feedback, which significantly improved our manuscript. Furthermore, we thank George Becker for sharing a few quasar spectra for a data comparison, as well as Vikram Khaire and Girish Kulkarni for helpful feedback and discussion. 

The data presented in this paper were obtained at the W.M. Keck
Observatory, which is operated as a scientific partnership among the
California Institute of Technology, the University of California and
the National Aeronautics and Space Administration. The Observatory was
made possible by the generous financial support of the W.M. Keck
Foundation.

This research has made use of the Keck Observatory Archive (KOA), which is operated by the W. M. Keck Observatory and the NASA Exoplanet Science Institute (NExScI), under contract with the National Aeronautics and Space Administration.

The authors wish to recognize and acknowledge the very significant cultural role and reverence that the summit of Mauna Kea has always had within the indigenous Hawaiian community.  We are most fortunate to have the opportunity to conduct observations from this mountain. 

Calculations presented in this paper used the hydra and draco clusters of the Max Planck Computing and Data Facility (MPCDF, formerly known as RZG). MPCDF is a competence center of the Max Planck Society located in Garching (Germany).

This work used the DiRAC Durham facility managed by the Institute for
Computational Cosmology on behalf of the STFC DiRAC HPC Facility
(\url{www.dirac.ac.uk}). The equipment was funded by BEIS capital funding
via STFC capital grants ST/P002293/1, ST/R002371/1 and ST/S002502/1,
Durham University and STFC operations grant ST/R000832/1. DiRAC is
part of the National e-Infrastructure.

\appendix

\section{Matching the Quasar Spectra to Photometry Measurements}\label{app:photometry}

In order to avoid biases in the optical depths measurements due to potential issues in the quasar spectra that could have been introduced in the fluxing or co-adding process, we verify that our final spectra match the observed photometric measurements in the $i$- and $z$-band, at a central wavelength of $7480$~{\AA} and $8932$~{\AA}, respectively. Thus we integrate the observed flux underneath the filter curves and compare these integrated magnitudes to the  photometry. To avoid contamination due to spurious negative pixels we clip all pixels that are lower than $2\sigma$. 

\begin{figure*}[t!]
\centering
\includegraphics[width=.9\textwidth]{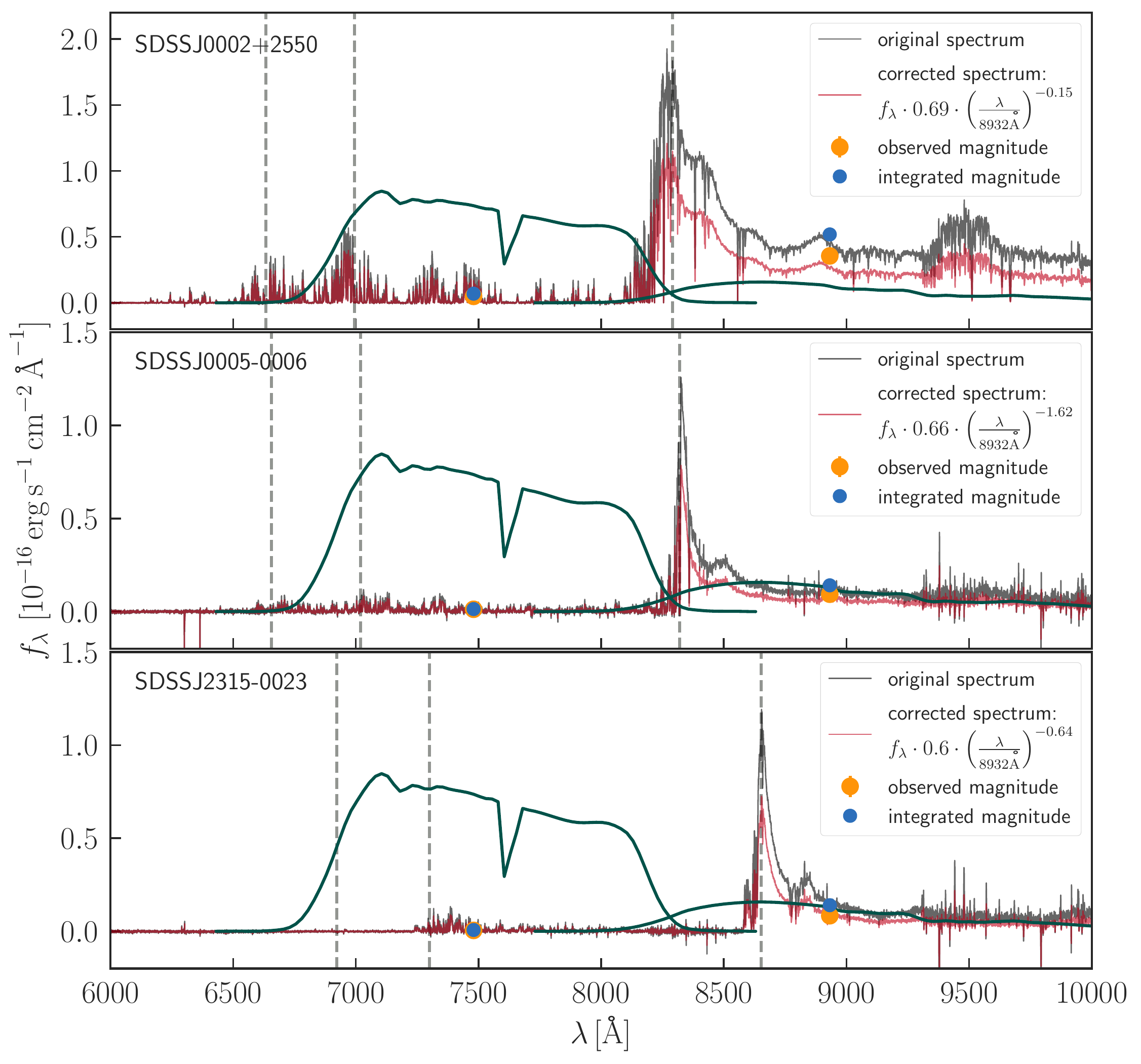}
\caption{Three examples of our slightly corrected spectra (red) once they are matched to the observed photometry (orange data point) in the $i$- and $z$-band. The original spectra are shown in black and the blue data points show the magnitudes in the $i$- and $z$-band from the original flux integrated underneath the SDSS filter curves (dark green). }\label{fig:examples_photometry}
\end{figure*}

If necessary we conduct a power-law correction to the quasar spectra: 
\begin{equation}
f_{\lambda, \rm\, new} = f_{\lambda} \cdot A \cdot \left(\frac{\lambda}{8932\rm {\AA}}\right)^{\alpha}, 
\end{equation}
by first calculating the amplitude $A$ by matching the $z$-band photometry, and afterwards estimating the slope $\alpha$ by requiring the $i$-band magnitudes to match as well. The boundaries for both $A$ and $\alpha$ were set to $[-3, 3]$. A few examples are shown in Fig.~\ref{fig:examples_photometry}. 

\section{Additional Figures and Tables}\label{app:chunks}

We show the remaining \lya\ optical depth measurements for all $40$~cMpc bins that do not have a \lyb\ forest measurement at the same absorption redshift $z_{\rm abs}$ in  Tab.~\ref{tab:measurements_lya}. 

\begin{figure*}[t!]
\centering
\includegraphics[width=\textwidth]{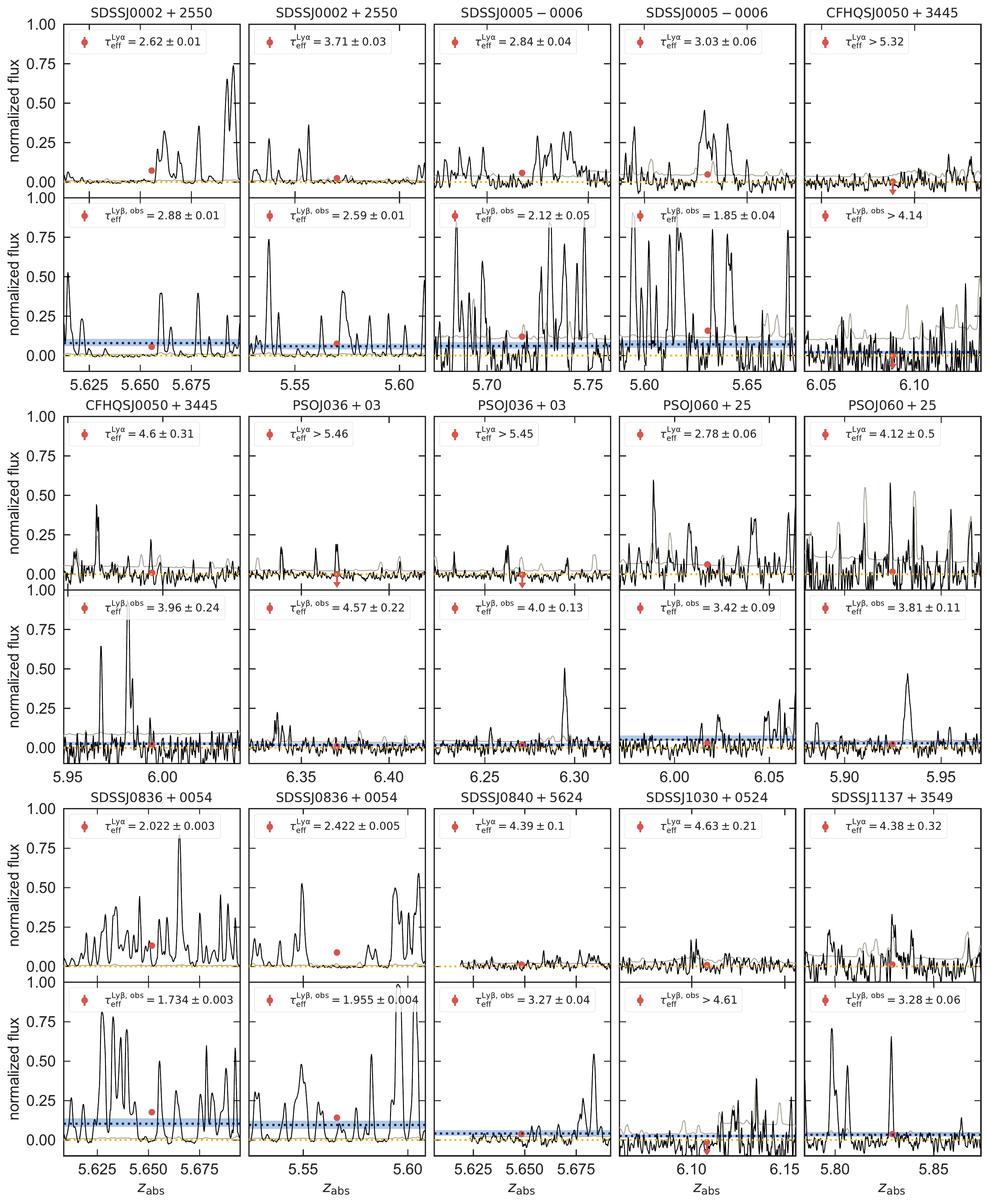}
\caption{Same as Fig.~\ref{fig:chunks} for the remaining spectral bins. }\label{fig:all_chunks1}
\vspace{.5 cm}
\end{figure*}

\begin{figure*}[t!]
\centering
\includegraphics[width=\textwidth]{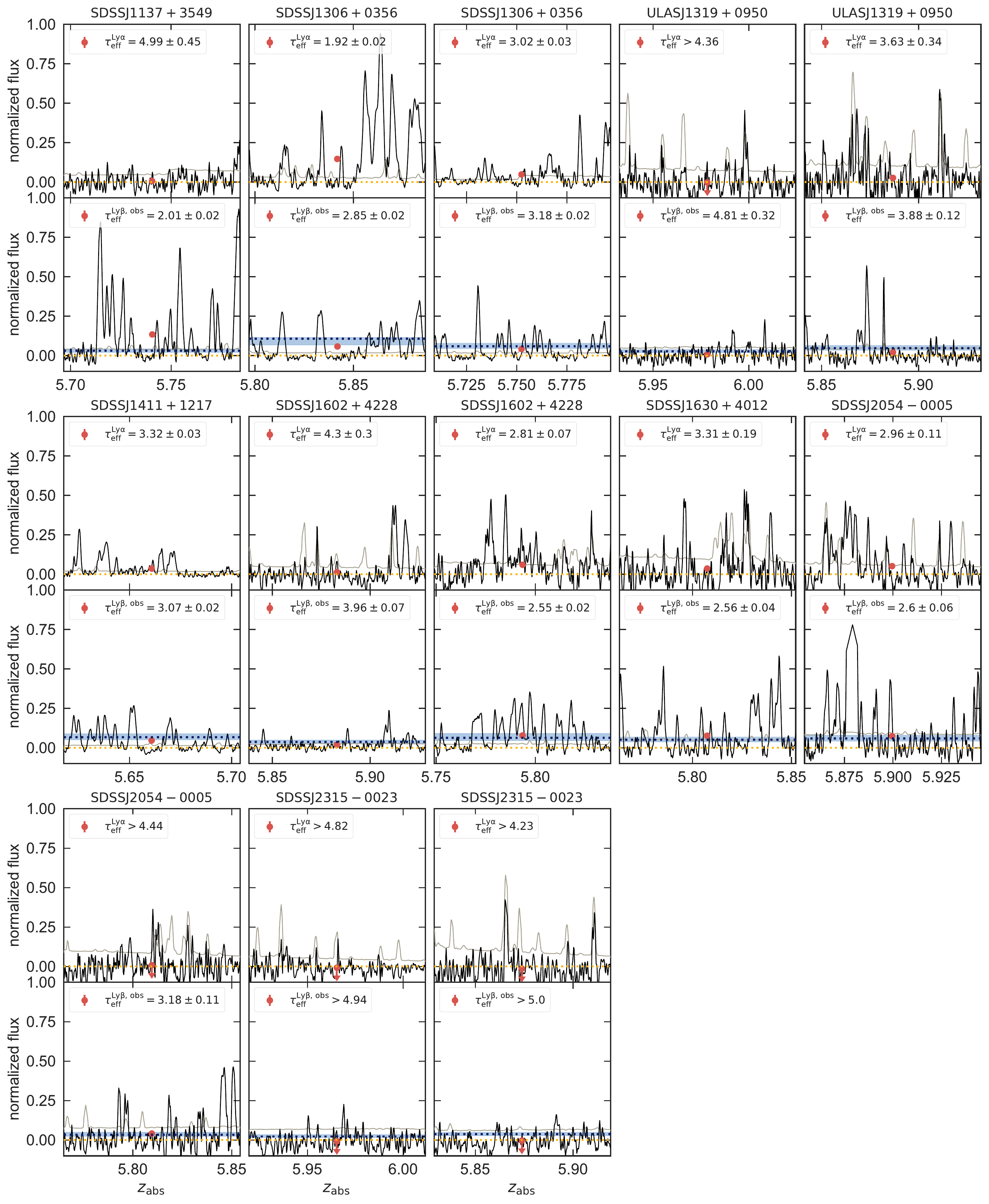}
\caption{Same as Fig.~\ref{fig:chunks} and Fig.~\ref{fig:all_chunks1}. }\label{fig:all_chunks2}
\vspace{.5 cm}
\end{figure*}

Fig.~\ref{fig:all_chunks1} and Fig.~\ref{fig:all_chunks2} show all spectral bins for which we have both measurements of $\tau_{\rm eff}^{\rm Ly\alpha}$ and $\tau_{\rm eff}^{\rm Ly\beta,\,obs}$ around the same absorption redshift.

\newpage
\startlongtable
\begin{deluxetable*}{lLLLLR}
\setlength{\tabcolsep}{15pt}
\tabletypesize{\footnotesize}
\tablecaption{Remaining mean flux measurements in the \lya\ forest. \label{tab:measurements_lya}}
\tablehead{\colhead{object} & \dcolhead{z_{\rm em}} & \dcolhead{z_{\rm start}} & \dcolhead{z_{\rm abs}} & \dcolhead{z_{\rm end}} & \dcolhead{\langle F^{\rm Ly\alpha}\rangle}}
\renewcommand{\arraystretch}{.9}
\startdata
SDSSJ0002+2550 & $5.82$ & $5.528$ & $5.486$ & $5.445$ & $0.0269\pm0.0007$ \\
 &  & $5.445$ & $5.404$ & $5.364$ & $0.0487\pm0.0007$ \\
 &  & $5.364$ & $5.324$ & $5.284$ & $0.0195\pm0.0006$ \\
 &  & $5.284$ & $5.244$ & $5.205$ & $0.0300\pm0.0005$ \\
 &  & $5.205$ & $5.166$ & $5.128$ & $0.1421\pm0.0007$ \\
 &  & $5.128$ & $5.090$ & $5.053$ & $0.0960\pm0.0008$ \\
 &  & $5.053$ & $5.015$ & $4.978$ & $0.1751\pm0.0009$ \\
 &  & $4.905$ & $4.869$ & $4.834$ & $0.0387\pm0.0004$ \\
SDSSJ0005-0006 & $5.844$ & $5.588$ & $5.545$ & $5.504$ & $0.0591\pm0.0043$ \\
 &  & $5.504$ & $5.462$ & $5.421$ & $0.0272\pm0.0046$ \\
 &  & $5.421$ & $5.380$ & $5.340$ & $0.0615\pm0.0052$ \\
 &  & $5.340$ & $5.300$ & $5.261$ & $0.1019\pm0.0037$ \\
 &  & $5.261$ & $5.221$ & $5.183$ & $0.0398\pm0.0045$ \\
 &  & $5.183$ & $5.144$ & $5.106$ & $0.0837\pm0.0046$ \\
 &  & $5.106$ & $5.068$ & $5.031$ & $0.1883\pm0.0059$ \\
 &  & $5.031$ & $4.994$ & $4.957$ & $0.1396\pm0.0063$ \\
 &  & $4.957$ & $4.920$ & $4.884$ & $0.0889\pm0.0036$ \\
 &  & $4.884$ & $4.848$ & $4.813$ & $0.1628\pm0.0022$ \\
CFHQSJ0050+3445 & $6.253$ & $5.948$ & $5.902$ & $5.857$ & $0.0395\pm0.0051$ \\
 &  & $5.857$ & $5.812$ & $5.767$ & $0.0275\pm0.0055$ \\
 &  & $5.767$ & $5.723$ & $5.680$ & $0.0456\pm0.0040$ \\
 &  & $5.680$ & $5.637$ & $5.594$ & $0.0420\pm0.0035$ \\
 &  & $5.594$ & $5.551$ & $5.509$ & $0.0762\pm0.0041$ \\
 &  & $5.509$ & $5.468$ & $5.427$ & $0.1910\pm0.0036$ \\
 &  & $5.427$ & $5.386$ & $5.346$ & $0.1997\pm0.0042$ \\
 &  & $5.346$ & $5.306$ & $5.266$ & $0.0767\pm0.0030$ \\
 &  & $5.266$ & $5.227$ & $5.188$ & $0.1952\pm0.0030$ \\
SDSSJ0100+2802 & $6.3258$ & $6.055$ & $6.008$ & $5.962$ & $0.0074\pm0.0005$ \\
 &  & $5.962$ & $5.916$ & $5.871$ & $0.0027\pm0.0007$ \\
 &  & $5.871$ & $5.826$ & $5.781$ & $0.0010\pm0.0007$ \\
 &  & $5.781$ & $5.737$ & $5.693$ & $0.0023\pm0.0005$ \\
 &  & $5.693$ & $5.650$ & $5.607$ & $0.0268\pm0.0004$ \\
 &  & $5.607$ & $5.565$ & $5.523$ & $0.0745\pm0.0006$ \\
 &  & $5.523$ & $5.481$ & $5.440$ & $0.0357\pm0.0005$ \\
 &  & $5.440$ & $5.399$ & $5.358$ & $0.1558\pm0.0005$ \\
 &  & $5.358$ & $5.318$ & $5.278$ & $0.1094\pm0.0005$ \\
ULASJ0148+0600 & $5.98$ & $5.666$ & $5.623$ & $5.580$ & $-0.0014\pm0.0025$ \\
 &  & $5.580$ & $5.538$ & $5.496$ & $0.0204\pm0.0032$ \\
 &  & $5.496$ & $5.455$ & $5.414$ & $0.0933\pm0.0030$ \\
 &  & $5.414$ & $5.373$ & $5.333$ & $0.0640\pm0.0031$ \\
 &  & $5.333$ & $5.293$ & $5.253$ & $0.0975\pm0.0020$ \\
 &  & $5.253$ & $5.214$ & $5.175$ & $0.1768\pm0.0025$ \\
 &  & $5.175$ & $5.137$ & $5.099$ & $0.1995\pm0.0026$ \\
 &  & $5.099$ & $5.061$ & $5.024$ & $0.2439\pm0.0038$ \\
 &  & $5.024$ & $4.987$ & $4.950$ & $0.1394\pm0.0029$ \\
PSOJ036+03 & $6.5412$ & $6.222$ & $6.173$ & $6.125$ & $-0.0003\pm0.0017$ \\
 &  & $6.125$ & $6.078$ & $6.031$ & $-0.0024\pm0.0014$ \\
 &  & $6.031$ & $5.984$ & $5.938$ & $-0.0024\pm0.0013$ \\
 &  & $5.938$ & $5.893$ & $5.847$ & $-0.0023\pm0.0018$ \\
 &  & $5.847$ & $5.802$ & $5.758$ & $0.0019\pm0.0012$ \\
 &  & $5.758$ & $5.714$ & $5.671$ & $0.0073\pm0.0007$ \\
 &  & $5.671$ & $5.628$ & $5.585$ & $0.0122\pm0.0009$ \\
 &  & $5.585$ & $5.542$ & $5.501$ & $0.0942\pm0.0023$ \\
 &  & $5.501$ & $5.459$ & $5.418$ & $0.0634\pm0.0021$ \\
PSOJ060+25 & $6.18$ & $5.879$ & $5.834$ & $5.789$ & $0.0163\pm0.0119$ \\
 &  & $5.789$ & $5.745$ & $5.701$ & $0.0390\pm0.0071$ \\
 &  & $5.701$ & $5.658$ & $5.615$ & $0.0041\pm0.0042$ \\
 &  & $5.615$ & $5.572$ & $5.530$ & $0.1139\pm0.0050$ \\
 &  & $5.530$ & $5.488$ & $5.447$ & $0.0321\pm0.0055$ \\
 &  & $5.447$ & $5.406$ & $5.366$ & $0.0196\pm0.0046$ \\
 &  & $5.366$ & $5.326$ & $5.286$ & $0.0721\pm0.0039$ \\
 &  & $5.286$ & $5.246$ & $5.207$ & $0.0649\pm0.0037$ \\
 &  & $5.207$ & $5.168$ & $5.130$ & $0.1850\pm0.0041$ \\
SDSSJ0836+0054 & $5.81$ & $5.524$ & $5.483$ & $5.441$ & $0.1431\pm0.0006$ \\
 &  & $5.441$ & $5.400$ & $5.360$ & $0.0754\pm0.0006$ \\
 &  & $5.360$ & $5.320$ & $5.280$ & $0.0452\pm0.0005$ \\
 &  & $5.280$ & $5.240$ & $5.201$ & $0.0909\pm0.0004$ \\
 &  & $5.201$ & $5.163$ & $5.124$ & $0.1142\pm0.0006$ \\
 &  & $5.124$ & $5.086$ & $5.049$ & $0.1535\pm0.0006$ \\
 &  & $5.049$ & $5.011$ & $4.975$ & $0.1599\pm0.0007$ \\
 &  & $4.975$ & $4.938$ & $4.902$ & $0.1314\pm0.0005$ \\
 &  & $4.902$ & $4.866$ & $4.830$ & $0.2934\pm0.0005$ \\
SDSSJ0840+5624 & $5.8441$ & $5.521$ & $5.480$ & $5.438$ & $0.0225\pm0.0018$ \\
 &  & $5.438$ & $5.397$ & $5.357$ & $0.1171\pm0.0016$ \\
 &  & $5.357$ & $5.317$ & $5.277$ & $0.0681\pm0.0013$ \\
 &  & $5.277$ & $5.237$ & $5.198$ & $0.0683\pm0.0013$ \\
 &  & $5.198$ & $5.160$ & $5.121$ & $0.1036\pm0.0016$ \\
 &  & $5.121$ & $5.083$ & $5.046$ & $0.1209\pm0.0018$ \\
 &  & $5.046$ & $5.008$ & $4.972$ & $0.1024\pm0.0021$ \\
 &  & $4.972$ & $4.935$ & $4.899$ & $0.1521\pm0.0013$ \\
 &  & $4.899$ & $4.863$ & $4.827$ & $0.1914\pm0.0010$ \\
SDSSJ1030+0524 & $6.309$ & $5.968$ & $5.922$ & $5.876$ & $0.0163\pm0.0037$ \\
 &  & $5.876$ & $5.831$ & $5.787$ & $0.0440\pm0.0041$ \\
 &  & $5.787$ & $5.743$ & $5.699$ & $0.0112\pm0.0022$ \\
 &  & $5.699$ & $5.655$ & $5.612$ & $0.0262\pm0.0017$ \\
 &  & $5.612$ & $5.570$ & $5.528$ & $0.1100\pm0.0025$ \\
 &  & $5.528$ & $5.486$ & $5.445$ & $0.0822\pm0.0028$ \\
 &  & $5.445$ & $5.404$ & $5.363$ & $0.1839\pm0.0028$ \\
 &  & $5.363$ & $5.323$ & $5.283$ & $0.1393\pm0.0023$ \\
 &  & $5.283$ & $5.244$ & $5.205$ & $0.1237\pm0.0016$ \\
SDSSJ1137+3549 & $6.03$ & $5.697$ & $5.653$ & $5.610$ & $0.1641\pm0.0026$ \\
 &  & $5.610$ & $5.568$ & $5.526$ & $0.0692\pm0.0031$ \\
 &  & $5.526$ & $5.484$ & $5.443$ & $0.1263\pm0.0034$ \\
 &  & $5.443$ & $5.402$ & $5.361$ & $0.1874\pm0.0033$ \\
 &  & $5.361$ & $5.321$ & $5.281$ & $0.1294\pm0.0028$ \\
 &  & $5.281$ & $5.242$ & $5.203$ & $0.2018\pm0.0025$ \\
 &  & $5.203$ & $5.164$ & $5.126$ & $0.1494\pm0.0031$ \\
 &  & $5.126$ & $5.088$ & $5.050$ & $0.0885\pm0.0038$ \\
SDSSJ1148+5251 & $6.4189$ & $5.998$ & $5.951$ & $5.906$ & $0.0006\pm0.0005$ \\
 &  & $5.906$ & $5.860$ & $5.815$ & $0.0022\pm0.0004$ \\
 &  & $5.815$ & $5.771$ & $5.727$ & $0.0078\pm0.0003$ \\
 &  & $5.727$ & $5.683$ & $5.640$ & $0.0160\pm0.0002$ \\
 &  & $5.640$ & $5.597$ & $5.555$ & $0.0166\pm0.0003$ \\
 &  & $5.555$ & $5.513$ & $5.471$ & $0.0154\pm0.0004$ \\
 &  & $5.471$ & $5.430$ & $5.389$ & $0.0419\pm0.0004$ \\
 &  & $5.389$ & $5.349$ & $5.309$ & $0.0566\pm0.0004$ \\
SDSSJ1306+0356 & $6.016$ & $5.709$ & $5.665$ & $5.622$ & $0.0673\pm0.0010$ \\
 &  & $5.622$ & $5.580$ & $5.537$ & $0.0408\pm0.0013$ \\
 &  & $5.537$ & $5.495$ & $5.454$ & $0.0347\pm0.0014$ \\
 &  & $5.454$ & $5.413$ & $5.372$ & $0.0902\pm0.0013$ \\
 &  & $5.372$ & $5.332$ & $5.292$ & $0.0666\pm0.0011$ \\
 &  & $5.292$ & $5.253$ & $5.214$ & $0.0620\pm0.0010$ \\
 &  & $5.214$ & $5.175$ & $5.136$ & $0.1032\pm0.0011$ \\
 &  & $5.136$ & $5.098$ & $5.060$ & $0.1284\pm0.0014$ \\
 &  & $5.060$ & $5.023$ & $4.986$ & $0.0801\pm0.0015$ \\
ULASJ1319+0950 & $6.133$ & $5.841$ & $5.796$ & $5.752$ & $0.0354\pm0.0087$ \\
 &  & $5.752$ & $5.708$ & $5.665$ & $0.0161\pm0.0040$ \\
 &  & $5.665$ & $5.622$ & $5.579$ & $0.0432\pm0.0041$ \\
 &  & $5.579$ & $5.537$ & $5.495$ & $0.0305\pm0.0054$ \\
 &  & $5.495$ & $5.454$ & $5.413$ & $0.0852\pm0.0058$ \\
 &  & $5.413$ & $5.372$ & $5.332$ & $0.0409\pm0.0049$ \\
 &  & $5.332$ & $5.292$ & $5.252$ & $0.0580\pm0.0034$ \\
 &  & $5.252$ & $5.213$ & $5.174$ & $0.1965\pm0.0045$ \\
 &  & $5.174$ & $5.136$ & $5.098$ & $0.1297\pm0.0045$ \\
SDSSJ1411+1217 & $5.904$ & $5.618$ & $5.575$ & $5.533$ & $0.0274\pm0.0016$ \\
 &  & $5.533$ & $5.491$ & $5.450$ & $0.0798\pm0.0019$ \\
 &  & $5.450$ & $5.409$ & $5.368$ & $0.0684\pm0.0016$ \\
 &  & $5.368$ & $5.328$ & $5.288$ & $0.0661\pm0.0014$ \\
 &  & $5.288$ & $5.249$ & $5.210$ & $0.0453\pm0.0016$ \\
 &  & $5.210$ & $5.171$ & $5.133$ & $0.1859\pm0.0017$ \\
 &  & $5.133$ & $5.095$ & $5.057$ & $0.1225\pm0.0019$ \\
 &  & $5.057$ & $5.019$ & $4.982$ & $0.1217\pm0.0024$ \\
 &  & $4.982$ & $4.946$ & $4.909$ & $-0.0032\pm0.0010$ \\
SDSSJ1602+4228 & $6.09$ & $5.749$ & $5.705$ & $5.662$ & $0.0218\pm0.0021$ \\
 &  & $5.662$ & $5.619$ & $5.576$ & $0.0583\pm0.0022$ \\
 &  & $5.576$ & $5.534$ & $5.492$ & $0.0660\pm0.0024$ \\
 &  & $5.492$ & $5.451$ & $5.410$ & $0.0441\pm0.0025$ \\
 &  & $5.410$ & $5.369$ & $5.329$ & $0.0679\pm0.0025$ \\
 &  & $5.329$ & $5.289$ & $5.249$ & $0.0633\pm0.0015$ \\
 &  & $5.249$ & $5.210$ & $5.171$ & $0.1517\pm0.0019$ \\
 &  & $5.171$ & $5.133$ & $5.095$ & $0.0625\pm0.0020$ \\
 &  & $5.095$ & $5.057$ & $5.020$ & $0.1376\pm0.0027$ \\
SDSSJ1630+4012 & $6.065$ & $5.763$ & $5.719$ & $5.675$ & $-0.0131\pm0.0033$ \\
 &  & $5.675$ & $5.632$ & $5.589$ & $0.0041\pm0.0030$ \\
 &  & $5.589$ & $5.547$ & $5.505$ & $0.0003\pm0.0038$ \\
 &  & $5.505$ & $5.463$ & $5.422$ & $0.1239\pm0.0043$ \\
 &  & $5.422$ & $5.381$ & $5.341$ & $0.0441\pm0.0046$ \\
 &  & $5.341$ & $5.301$ & $5.262$ & $0.1392\pm0.0028$ \\
 &  & $5.262$ & $5.222$ & $5.184$ & $0.0906\pm0.0037$ \\
 &  & $5.184$ & $5.145$ & $5.107$ & $0.1107\pm0.0039$ \\
 &  & $5.107$ & $5.069$ & $5.032$ & $0.1986\pm0.0048$ \\
SDSSJ2054-0005 & $6.0391$ & $5.765$ & $5.721$ & $5.678$ & $0.0179\pm0.0043$ \\
 &  & $5.678$ & $5.635$ & $5.592$ & $0.0530\pm0.0045$ \\
 &  & $5.592$ & $5.549$ & $5.507$ & $0.1882\pm0.0062$ \\
 &  & $5.507$ & $5.466$ & $5.425$ & $0.1527\pm0.0062$ \\
 &  & $5.425$ & $5.384$ & $5.344$ & $0.0782\pm0.0063$ \\
 &  & $5.344$ & $5.304$ & $5.264$ & $0.1855\pm0.0042$ \\
 &  & $5.264$ & $5.225$ & $5.186$ & $0.1197\pm0.0053$ \\
 &  & $5.186$ & $5.148$ & $5.110$ & $0.2639\pm0.0067$ \\
 &  & $5.110$ & $5.072$ & $5.034$ & $0.1857\pm0.0095$ \\
SDSSJ2315-0023 & $6.117$ & $5.740$ & $5.696$ & $5.653$ & $0.0261\pm0.0048$ \\
 &  & $5.653$ & $5.610$ & $5.567$ & $0.0708\pm0.0058$ \\
 &  & $5.567$ & $5.525$ & $5.483$ & $0.0276\pm0.0063$ \\
 &  & $5.483$ & $5.442$ & $5.401$ & $0.0390\pm0.0067$ \\
 &  & $5.401$ & $5.360$ & $5.320$ & $0.0762\pm0.0054$ \\
 &  & $5.320$ & $5.280$ & $5.241$ & $0.1411\pm0.0044$ \\
 &  & $5.241$ & $5.202$ & $5.163$ & $0.0920\pm0.0051$ \\
 &  & $5.163$ & $5.125$ & $5.087$ & $0.2344\pm0.0045$ \\
 \enddata
\tablecomments{The different columns show the name of the object and its emission redshift $z_{\rm em}$, the start of the redshift bin $z_{\rm start}$, the mean redshift of each bin $z_{\rm abs}$, and the end of each bin $z_{\rm end}$, as well as the measured mean flux of the continuum normalized spectra in the \lya\ forest. }
\end{deluxetable*}

\section{Numerical Convergence of the Hydrodynamical Simulation}\label{app:resolution}

In this section we discuss the numerical convergence of the the \lyb\ forest in the \texttt{Nyx} hydrodynamical simulation. While several studies have shown the convergence of different \lya\ forest statistics \citep[see e.g. Fig.~$6$ in][]{Onorbe2017, Bolton2017}, there has not been any convergence test on the \lyb\ forest. We present here a first study on the \lyb\ convergence making use of a publicly available suite of more than $60$ \texttt{Nyx} hydrodynamical simulations\footnote{THERMAL suite: \url{http://thermal.joseonorbe.com/}}. 
We select several simulations from this suite in which we only changed the spatial resolution or the box size. Thus, all simulations use exactly the same photoionization rates at all
redshifts. 

\begin{figure*}[t!]
\centering
\includegraphics[width=\textwidth]{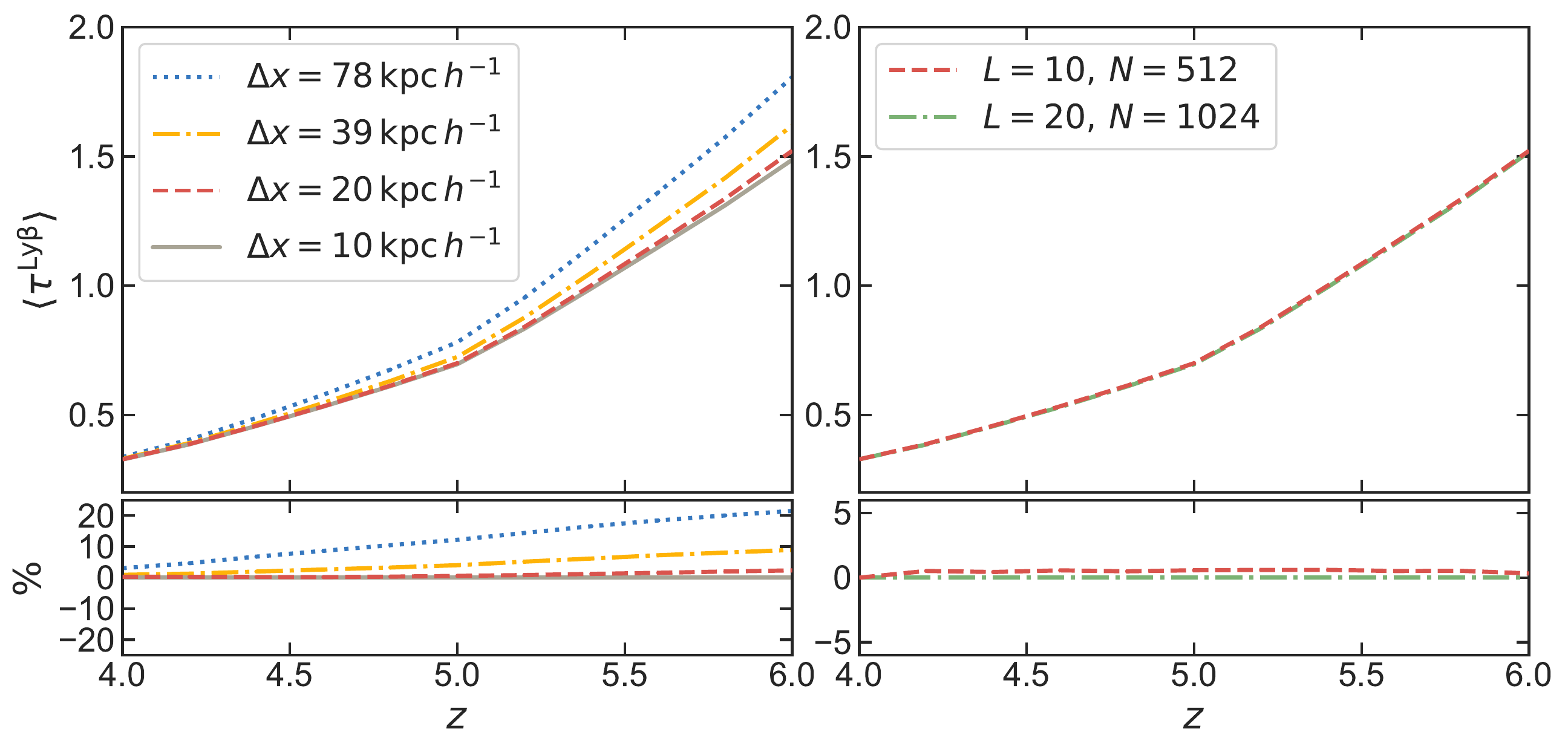}
\caption{Convergence of the mean \lyb\ optical depth $\langle \tau^{\rm Ly\beta}\rangle$ at $4.0\leq z\leq 6.0$. \textit{Left panel:} simulations with a fixed box size ($L=10$ Mpc h$^{-1}$) and different spatial
resolution, i.e. $\Delta x=78\,{\rm kpc}\,h^{-1}$ (blue dotted), $39\,{\rm kpc}\,h^{-1}$ (yellow dash-dotted), $20\,{\rm kpc}\,h^{-1}$ (red dashed), and
$10\,{\rm kpc}\,h^{-1}$ (grey). \textit{Right panel:} simulations with a fixed
spatial resolution ($\Delta x\sim 20\,{\rm kpc}\,h^{-1}$) and different box sizes, i.e. $L=10\,{\rm Mpc}\,h^{-1}$ (red dashed), $20\,{\rm Mpc}\,h^{-1}$ (green dash-dotted).}\label{fig:resolution_lyb}
\end{figure*}

We compute the \lyb\ mean optical depth evolution directly from the simulated mean flux using Eqn.~\ref{eq:tau}, without any rescaling of the photoionization rate. Convergence tests for the high-$z$ \lya\ forest using the same suite of simulations can be found in \citet{Onorbe2017}, and we refer the reader to their work for further details on the simulations. 
The left panel of Figure~\ref{fig:resolution_lyb} shows the evolution of the mean \lyb\ optical depth for four simulations with the same box size, i.e.  $L=10\,{\rm Mpc}\,h^{-1}$, but increasing numbers of resolution elements: $128^3$ (blue dashed), $256^3$ (yellow dash-dotted), $512^3$ (red dashed), and $1024^3$ (grey), which results in cell sizes of $\Delta x = 78$, $39$, $20$, and
$10\, {\rm kpc }\,h^{-1}$, respectively. The $512^3$ run has the same spatial resolution as the models discussed in this work (see \S~\ref{sec:sims}). 
The right panel shows the convergence for different box sizes, i.e. $L=10\,{\rm Mpc}\,h^{-1}$ and $20\,{\rm Mpc}\,h^{-1}$, at the same spatial resolution ($\Delta x = 20\, {\rm kpc }\,h^{-1}$). 
The simulations discussed in this work (red dashed lines in both panels) show a good convergence level below  $<5\%$ between $4\leqslant z\leqslant6$, both in terms of spatial resolution as well as box size. Therefore we do not expect any significant effect due to the resolution and/or box size of the simulations on the main conclusions of this work. 

\software{\href{http://www2.keck.hawaii.edu/inst/esi/ESIRedux/}{ESIRedux}, \href{http://www.ucolick.org/~xavier/IDL/}{XIDL}, \href{http://specdb.readthedocs.io/en/latest/igmspec.html}{igmspec}, numpy \citep{numpy}, scipy \citep{scipy}, matplotlib \citep{matplotlib}, astropy \citep{astropy}}

\bibliography{literatur_hz}

\end{document}